\documentclass[%
 reprint,
 amsmath,amssymb,
 aps,
]{revtex4-1}

\usepackage{graphicx}% Include figure files
\usepackage{amsmath}
\usepackage{dcolumn}% Align table columns on decimal point
\usepackage{bm}% bold math
\usepackage{color}
\usepackage[normalem]{ulem}
\usepackage{url}
\begin{document}

    \title{Intrinsic structure perspective for MIPS interfaces in two dimensional  systems of Active Brownian Particles}% Force line breaks with \\
    %\thanks{A footnote to the article title}%
    
    \author{Enrique Chacon}
    \affiliation{Instituto de Ciencia de Materiales de Madrid, CSIC, 28049 Madrid, Spain}
    \author{Francisco Alarcón}
    \affiliation{Departamento de Estructura de la Materia, F\'isica T\'ermica y Electr\'onica, Facultad de Ciencias F\'isicas, Universidad Complutense de Madrid, 28040, Madrid, Spain}
    \affiliation{Departamento de Ingeniería Física, División de Ciencias e Ingenierías, Universidad de Guanajuato, Loma del Bosque 103, 37150 León, Mexico.}
 \author{Jorge Ram{\'i}rez}
\affiliation{Dep. de Ingenier{\'i}a Qu{\'i}mica, ETSI Industriales, Universidad Polit{\'e}cnica de Madrid, 28006 Madrid, Spain.}
     \author{Pedro Tarazona}
    \affiliation{Departamento de Física Teórica de la Materia Condensada, Condensed Matter Physics Center (IFIMAC), Universidad Autonoma de Madrid, 28049 Madrid, Spain}

    \author{Chantal Valeriani}%
    \affiliation{Departamento de Estructura de la Materia, F\'isica T\'ermica y Electr\'onica, Facultad de Ciencias F\'isicas, Universidad Complutense de Madrid, 28040, Madrid, Spain}
    \email{cvaleriani@ucm.es}

    \date{\today}% It is always \today, today,
                 %  but any date may be explicitly specified
    
    \begin{abstract}
   Suspensions of  Active Brownian Particles (ABP) undergo motility induced phase separation (MIPS) 
over a wide range of mean density and activity strength  \cite{Stenhammar2014}, which
implies the spontaneous aggregation of particles due to the persistence of their direction of
motion, even in the absence of an explicit attraction. 
Both similarities and qualitative differences have been obtained when the  MIPS is analysed in the same terms as a liquid-gas 
phase coexistence in an equilibrium attractive system.  Negative values of the mechanical surface tension have been reported,  
from the total forces across the interface, while the stable fluctuations of the interfacial line would be interpreted as a positive 
capillary surface tension \cite{PRLSpeck}, while in equilibrium liquid surfaces these two magnitudes are equal. 
 We present here the analysis of 2D-ABP interfaces in terms of the intrinsic density and force profiles, calculated with 
 the particle distance to the instantaneous interfacial line. Our results provide a new insight in the origin of the MIPS from the 
 local rectification of the random active force on the particles near the interface. As it had  been pointed, that effect acts as an 
 external potential \cite{Omar_PRE_2020} that produces a pressure gradient  across the interface, so that the mechanical surface tension 
 of the MIPS cannot be described as that of equilibrium coexisting phases; but our analysis shows that most of that effect 
 comes from the tightly caged particles at the dense (inner) side of  the MIPS interface, rather than from the free moving 
 particles at the outer side that collide with the dense cluster. Moreover, a clear correlation appears between the decay of 
 the hexatic order parameter at the dense slab and the end of the MIPS as the strength of the active force is lowered. 
 We test that with the strong active forces required for MIPS, the interfacial structure and properties are very similar 
 for ABP with purely repulsive (WCA-LJ model truncated at its minimum)  and when the interaction includes a range 
 of the LJ attractive force.  
     
    \end{abstract}

\maketitle

\section{\label{Intro}Introduction}

Active matter focuses on systems  composed of particles capable of  consuming energy in order to move, constantly
driving themselves away from equilibrium\cite{Bechinger2016}.
Characteristic features of a bulk suspension of active particles  are collective  behaviours such as large scale directed motion, dynamic clusters formation or non-equilibrium phase separation. 
A numerical model that has been thoroughly exploited to understand such properties consists of  a suspension of Active Brownian Particles (ABP). ABP are self-propelled Brownian
particles whose inter-particle interactions can be either  purely repulsive\cite{Stenhammar2014,RednerPRL} 
or contain also some range of attractive force\cite{Redner_aabp,Mognetti2013,AlarconSoftmatter}.

Depending on the nature of  inter-particle interactions and of the particles' activity, the system forms different steady state structures. 
On the one side, a dilute suspension of attractive ABP is characterised by a distribution of dynamic clusters whenever the inter-particle attraction strength is enough to compete with the particles' activity    \cite{Redner_aabp,Mognetti2013,Sarkar2021,AlarconSoftmatter}, without any phase separation taking place. 
Clustering is due to a balance between new particles joining already formed clusters  and large clusters breaking  into smaller ones \cite{palacci,Buttinoni2013,Theurkauff}.
On the other side, a suspension of repulsive ABP undergoes motility induced phase separation (MIPS) for relatively high densities and activity strength  \cite{Stenhammar2014,Hermann2021}.
MIPS implies a spontaneous aggregation of particles due to the persistence of their direction of
motion, even in the absence of an explicit attraction.
MIPS has been detected  in experimental set-ups  of active colloids  \cite{Buttinoni2013,Thutupalli2011,Sano2015,DauchotPRL2016} and thoroughly characterised  in numerical simulations of Active Brownian particles (ABP)\cite{Stenhammar2014,RednerPRL,DiGregorio_PRL_2018,JoseMIPS,SolonNJP,Bettolo2019,SoftMatterPagonabarragaa,SolonPRL2020,Wysocki2014}. Moreover, MIPS cannot be avoided in the presence of    inertial effects  \cite{JCPLowen2020,Hou2021}, when considering a suspension of active particles moving on a lattice \cite{MIPSvirnau,barri2021}, or when an amount of passive particles is added to the active suspension   \cite{DiegoMix2020,StenhammarPRL,Wysocki2016}.

Trying to understand this non-equilibrium phase separation via a mechanical equation of state, one could draw similarities with the interfacial properties of equilibrium phases. However, a clear  agreement has not been reached on the physical properties such as the surface tension 
%Even though the  dense-dilute MIPS interface  is largely fluctuating
\cite{MIPSBand_BinderCum, Patch_SoftMat_2018,Omar_PRE_2020,JCPdijkstra2017,Nardini_2021,Ginot2015}.
The main reason behind the disagreement can be summarised in the difference between considering  activity giving rise to
a spatially uniform stress 
%that acts as  a spatially uniform state variable 
(the swim
pressure \cite{Fily2014,Takatori2014,mallory2020}) or activity giving rise to a spatially varying body force\cite{Epstein2019,Omar_PRE_2020}.

Even though the swim pressure has been  successfully used to determine the onset of MIPS \cite{SoftMatterPagonabarragaa,Winkler2015,Patch2017,SolonNatPhys,   Speck2016,Fily2018,Das2019}, 
when dealing with localised phenomena, adding the swim pressure to the   total pressure results in  a negative surface tension, that
in equilibrium phase coexistence would imply an unstable interface.  
Considering a "stable" (even though largely fluctuating) dense-dilute MIPS slab configurations in  a non-square box\cite{MIPSBand_BinderCum,PRLSpeck}, it has been shown that  \cite{PRLSpeck}
the pressure formulation,  combining  the swim pressure  with  the Kirkwood and Buff formulation for the surface tension, leads to 
a negative mechanical interfacial tension compatible  with  a positive surface stiffness, or capillary surface tension. 
This is a strongly non-equilibrium effect, since from coexisting phases in a passive 
attractive system the mechanical and capillary surface tensions are equal \cite{Alejandre1999}.

To reconcile a stable interface with negative values of surface tension, the authors of Ref.\cite{Patch_SoftMat_2018} discovered a
strong correlation between the local curvature of the fluctuating interface and the magnitude of the surface tension. 
The local curvature is responsible for 
a local tangential motion of particles within a surface layer both in the dense and dilute phase:  the combined particles' motion in the dense and dilute phase leads to a stiffening interface that  reduces the amount of  local fluctuations.

Another way to reconcile  a stable interface with a negative values of the surface tension consists in not including  the swim pressure  in the stress calculations, as in Ref.\cite{PRLSpeck}. Instead, one can consider 
the effect of particle's activity via a body force, generated by the net alignment of ABP
at the interface \cite{Omar_PRE_2020}. As a result, this approach  leads to
a negligible surface tension in the dense–dilute interface of phase separated ABP \cite{SoftMatterKlotsa}.
The crucial difference with the other perspective is that, since the activity of the particles creates a net mean force
at the interfaces, the two MIPS phases  are like equilibrium regions with different potential energies, and therefore
different bulk pressures, rather than being similar to coexisting phases at equilibrium. The mechanical surface tension across
a (smooth) external potential step, that separates regions with different pressure,  may be negative in equilibrium systems, and it 
certainly differs from the  (always positive) capillary surface tension extracted from the fluctuations of that (externally induced) interface. 

Therefore, a key point to understand the physics of MIPS is the origin and structure of that self-sustained effective external potential
formed at the interfaces.  A major issue present when  extracting information on that effect from ABP simulations is that, at any fixed point $(x,y)$ 
located on the interfacial region,  the strong interfacial fluctuations average both density and force contributions of instantaneous configurations: this affects the point's local environments,  
%that happen to put on that point very different local environments, 
mixing low density and high density contributions, from both 
%one and the other side 
sides of the instantaneous boundary of the dense cluster. The same problem appears in molecular dynamics simulations
of equilibrium liquid surfaces, giving rise to density profiles that become  smoother for larger sizes, without a thermodynamic limit.

The Capillary Wave Theory (CWT)\cite{Rowlinson1982,Buff_CWT_1965} formalizes the account of these interfacial fluctuations and, although some of its
hypothesis (like the equivalence between the macroscopic and the capillary surface tensions) become problematic for  
active  systems away from equilibrium \cite{JCPdijkstra2017,Omar_PRE_2020}, it has already been used as a generic 
mesoscopic framework to describe the fluctuating MIPS interfaces \cite{Patch_SoftMat_2018}. In particular, we take advantage of  the methods developed to establish  a quantitative link between the CWT description of the interfaces as smooth fluctuations shapes, 
and the particle positions in computer simulations \cite{chacon_ISM_2003_PRL,HIDRODYNAMIC_ISM_PRL_2008,bolas}, both for equilibrium and dynamical properties.

In the present work,  we extend 
that analysis, with emphasis on obtaining intrinsic profiles for the density and for the active force acting on the particles. 
These intrinsic profiles give a sharper view for the formation of  MIPS, with the focus on the inner side
of the interface, where the particles are already quite tightly caged by their neighbours in local hexatic order structures.
The decay of the capillary tension towards the end of the MIPS, as the active force is lowered, is correlated with the
gradual lost of the global hexatic order at the inner part of the dense slab.  The authors of Ref.\cite{Nardini_2021} have recently 
demonstrated, for a minimum continuum active model, how  a negative  capillary tension may result into a microphase-separated 
state or into an active foam state, rather than the MIPS of a large dense cluster. 
 
 %We compare the results for ABP with WCA-LJ repulsive interactions with those including an attractive range ($r_c=2.5 \sigma$)
% from the tail of the LJ potential. Over the range of high activity strength that produces MIPS we find no qualitative difference
% and very little quantitative effects of that attractive interaction, that becomes crucial for much lower strength  of the active force
% when a equilibrium-like condensation phase transition may appear. 

We propose to make use of the approach presented in  Ref.\cite{Tarazona_CW_2007_PRL},  based on a  density functional description for capillary wave fluctuations on
free liquid surfaces. The reason behind this choice is that this approach can be safely applied to non-equilibrium systems. 
CWT can be applied to a liquid or a solid surface, as long as the system is kept above the roughening temperature. 
Our system satisfies this condition, being two dimensional and active. Moreover, choosing to study the surface fluctuations at small wavevectors guarantees to safely apply CWT, given the low upper-cutoff wavevector typical of a solid surface.

Thus, we estimate the surface tension in a phase separated system of Active Brownian Particles. 
We unravel the relevance of inter-particle interactions (whether repulsive or attractive) in both stability and 
structural features of the dense MIPS phase. 
We compare the results for ABP with WCA-LJ repulsive interactions with those including an attractive range ($r_c=2.5 \sigma$)
 from the tail of the LJ potential. Over the range of high activity strength that produces MIPS we find no qualitative difference
 and very little quantitative effects of that attractive interaction, that becomes crucial for much lower strength  of the active force
 when a equilibrium-like condensation phase transition may appear.

\section{Numerical and theoretical details}

\subsection{\label{sec:SimDetails}Model and Numerical details}

A suspension of active Brownian particles (ABP) consists of 
 particles undergoing a Brownian motion with an additional constant self-propelling force $F_a$ acting along their orientation vector $\vec{n}=(\cos \theta_i,\sin \theta_i)$, that fluctuates independently for each particle. The equations of motion are
\begin{align}
\label{eq:motion}
& \dot{\vec{r}}_i = \frac{D}{k_B T} \left( - \sum_{j\neq i} \nabla V(r_{ij}) + | F_a |\, \vec{n}_i \right) + \sqrt{2D} \, \vec{\xi}_i, 
\end{align}
\begin{align}
& \dot{\theta}_i = \sqrt{2D_r}\, \eta_i.
\label{eq:orient}
\end{align}
being $\eta_i$ and the components of $\vec{\xi}_i = (\xi^{(1)}_i(t), \xi^{(2)}_i(t))$ a white  stochastic noise with zero mean and 
correlations  $\langle \eta_i(t)\eta_j(t') \rangle = \langle \xi^{(1)}_i(t)\xi^{(1)}_j(t')  \rangle = \langle \xi^{(2)}_i(t)\xi^{(2)}_j(t')  \rangle =\delta_{ij} \delta(t-t')$.
$V(r_{ij})$ is the inter-particle pair   potential: to study a suspension of repulsive ABP, we consider a  WCA, i.e. a LJ truncated and shifted at its minimum $r_c=2^{1/6}\sigma$; whereas for a suspension of attractive ABP, we consider a LJ truncated and shifted at $r_c=2.5 \sigma$, that includes attractive interactions. 
%in our model a truncated Lennard-Jones (LJ), with 
%As units of length and energy, we the usual parameters as  units of lenght 
$\sigma=1$ and $\epsilon=1$  are units of length and energy. 
%We compare results with
%the WCA repulsive interaction, i.e. the LJ truncated and shifted at its minimum $r_c=2^{1/6}\sigma$, and for truncation at
%$r_c=2.5 \sigma$, that includes attractive %interactions.  
The temperature $T$, in units of the Boltzmann constant $k_B$ and 
$\epsilon$, is set at $k_B T/\epsilon=0.42$. We use the translational diffusion constant $D=1$ to set the time scale, $\tau=\sigma^2/D$. The rotational diffusion  is set to obey the Stokes-Einstein equations: $D_r = 3D/\sigma^2$.

 We  carry out simulations of a two dimensional suspension of ABP with an in house modified version of the {\it LAMMPS} \cite{LAMMPS,Diego_SoftMat2020}  open source package. The system consists of $N = 40000$  particles  in a two-dimensional box with periodic boundary conditions, that may be kept with fixed $(N,V,T)$
 or fluctuating $(N,p,T)$ to stabilize and analyse a MIPS slab (see Supplementary Info for more details).  The time step  has been set to $\Delta t = 10^{-5} \tau$.

 As a measure of the degree of  activity, we define the  dimensionless $F_a \sigma/\epsilon$, i.e. the constant active force modulus $F_a$ 
  in units of the LJ parameters \footnote{Note that this ratio is related to the $\xi$ reported in Ref\cite{AlarconSoftmatter}}.
 %  However, given our choice of $\sigma$ and $\epsilon$, we use   the constant active force modulus $F_a$ 
  %may be used as dimensionless number ($F_a \sigma/\epsilon$)  in units of the LJ parameters.
 Having to compare repulsive versus attractive particles,  
for large values of $F_a$, 
and for the structural properties explored  here, 
the value of the LJ-scaled $F_a$ is more relevant that the thermal   P\'eclet number
%\begin{equation}
$\text{Pe} =  \frac{3v}{\sigma D_r}$,
%\end{equation}
where $v=|F_a|D/k_BT$ is the self-propelling velocity\cite{Stenhammar2014}. 
%{\bf PACO:} \textit{I'm agree with this. $F_a$ as parameter of activity instead of $\xi$. }
Using the above mentioned parameters, we can relate  the P\'eclet number $\text{Pe} =  \frac{3v}{\sigma D_r}$ to the Peclet obtained by means of our parameter choice: 
$\text{Pe} =  \frac{3v}{\sigma D_r} = \frac{F_a \sigma}{\epsilon} \frac{1}{0.42}$.

%{\bf Paco y Chantal,  mirad si esto esta bien dicho y si quereis cambiar en todo el articulo la variable Fa por vuestra $\xi$...pero puede ser un lio ua que es el mismo simbolo que usamos para la superficie intrinseca, creemos que lo mejor es dejarlo como esta advirtiendo que Fa es igual qu vuestra $\xi$}

\subsection{\label{sec:theory}Theoretical details}

The Capillary Wave Theory~\cite{Tarazona_CW_2007_PRL} (CWT) describes the fluctuations of an interface with surface tension $\gamma_o$
through an {\sl intrinsic density profile} (IDP) $\rho_{I}(x)$ that follows the instantaneous shape of an {\sl intrinsic surface} (IS). In our 2D system
the interface is really  a line $x=\xi(y)=\sum_q \hat{\xi}_q e^{i q y}$  (considering the $Y$ axis along the dense slab), but we keep the trend
of previous authors and use "intrinsic surface" and "surface tension", instead of "intrinsic line" and "line tension".  Nevertheless, the 1D dimensional
character of the interface is reflected in an enhanced dependence of the mean density profile $\rho(x)$ with the size of the system.

The CWT describes that effect from the average over the fluctuations of the IS Fourier components $\hat{\xi}_q$: $\rho(x)=\langle \rho_{I}(x-\xi(y))\rangle_\xi$, which in a thermal equilibrium system could be predicted from Bolztmann distribution.  In our  ABP system we cannot use
such thermal equibrium hypothesis, but still we may cast in the CWT terms of the IS and the IDP the results of our simulations.
%. Thus, CWT  allows  to know the intrinsic profile of the dilute-dense  interface, and the IS fluctuations. 
\begin{figure}[h!]
\centering
\includegraphics[width=0.6\linewidth]{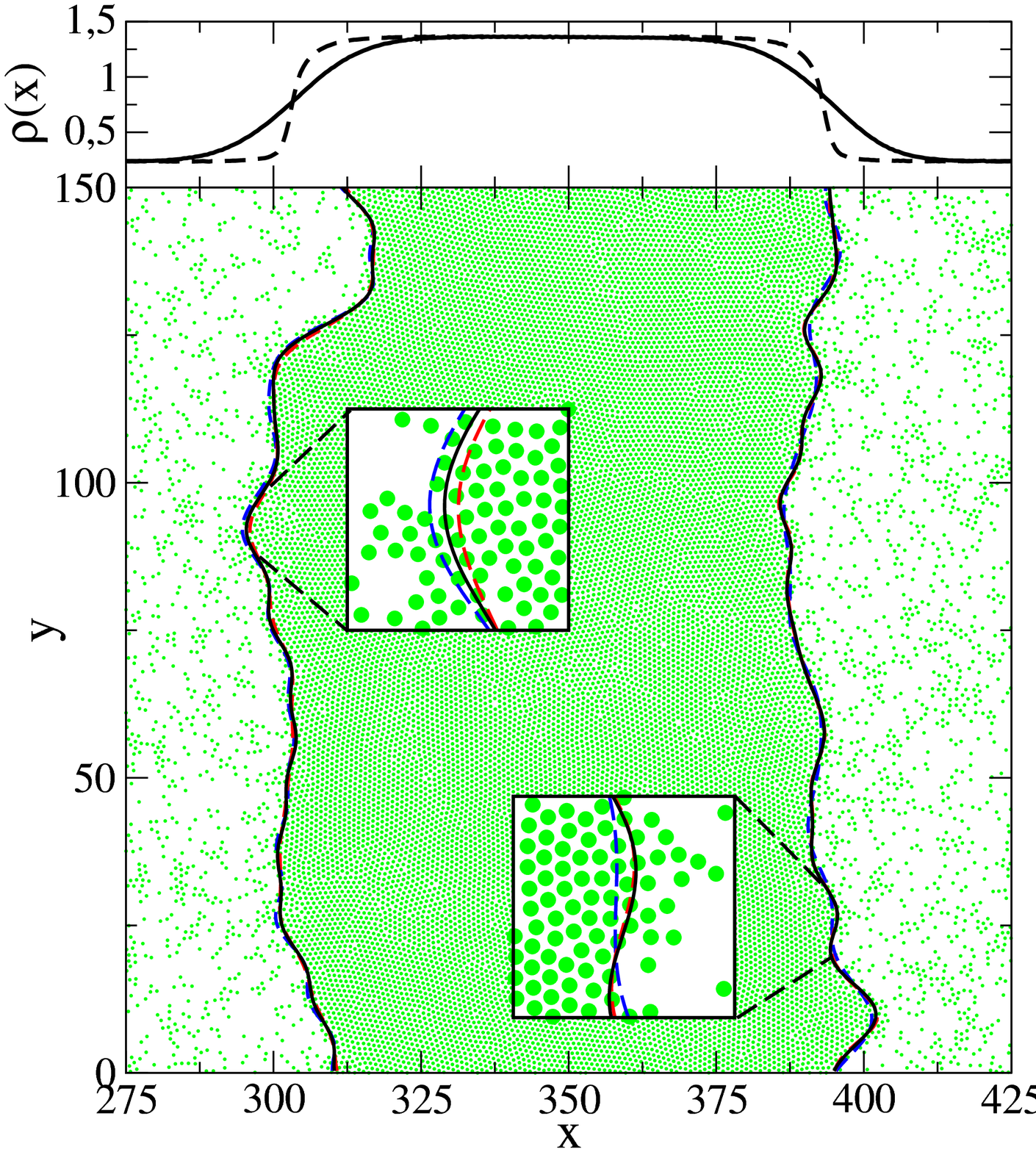}
\caption{Snapshot of the attractive ABP system, characterised by  $F_a=55$.   
 The IS is drawn at the two edges of the dense slab, for different choices of the parameters. The black lines (on both sides) correspond to our preferred choice, $\alpha=2$ for the Gaussian coarse-grain and $q_u=0.5$ as the upper cutoff in the Fourier series. On the left side-inset, the dashed lines show the effect of changing to $\alpha=1$ (red) and $\alpha=3$ (blue), keeping   $q_u=0.5$. On the right side-inset, we keep $\alpha=2$ and change to $q_u=0.25$ (red) and $q_u=1.5$ (blue). 
%The insets magnify to small regions to make visiblei the differences. 
Top panel:  mean density profile $\rho(x)$ (continuous line);  intrinsic  density profiles $\rho_{\hbox{\tiny I}}(x)$ (dashed line), with their origin  set at the mean position of each edge.  %We use LJ $\sigma=1$ as unit of length all over.
}
\label{fig:Snapshot}
\end{figure}

Considering an ABP suspension undergoing MIPS (Fig.~\ref{fig:Snapshot}),   a density gap
$\delta\rho_{hl}=\rho_l-\rho_h$ appears between the
low and high density phases. 
 Thus, we use a Gaussian smoothed density distribution $\varrho(x,y)= \frac{1}{2 \pi \alpha}\sum_{i} e^{- (\vec{r}-\vec{r}_i)^2/(2 \alpha^2)}$,
being $\vec{r}_i$ particle positions, to 
define the IS from the equation $\varrho(\xi(y),y)=(\rho_h+\rho_l)/2$, for each $y$, and to get the Fourier components $\hat{\xi}_q$ for the wavevectors allowed by the cell size $L_y$, up to an upper cutoff $|q|\leq q_{u}$.
Fig.~\ref{fig:Snapshot} represents a typical MIPS configuration (bottom panel, for an attractive ABP system) together with a density profile (top panel). The snapshot in the bottom panel  illustrates ABP (in green) and the   $x=\xi(y)$ lines  on both sides of the dense phase boundaries  (in black), calculated  with our choice $\alpha=2 \sigma$ and $q_{u}=0.5/\sigma$. 
To estimate the resolution of our mesoscopic description, we 
%The two insets 
report in the two insets the  $x=\xi(y)$ lines for a different choice of  $\alpha$ and $q_{u}$. 
%, to estimate the resolution of our mesoscopic description. 
%The colour dashed lines show other choices for $\alpha$ and $q_{u}$, their differences are visible in the magnified insets, that give a fair idea for the resolution level of the mesoscopic CWT descrition.
The top panel in Fig.~\ref{fig:Snapshot} shows the mean density profile $\rho(z)=\langle \sum_i \delta(x-x_i)\rangle/L_y$ that averages the position of the particles over 2000 configurations separated by 100000 steps.

As suggested by Ref.~\cite{Omar_PRE_2020}, in an ABP-MIPS system, despite the random direction $\vec{n}_i$ of the active force on each particle,
the mean active force profile $\vec{f}^{a}(x)=F_{a} \langle \sum_i \vec{n}_{i}\delta(x-x_{i}) \rangle /L_{y}$ does not vanish at the interface. Therefore, 
as for the IDP, we may get the intrinsic active force profiles $\vec{f}^{a}_{\hbox{\tiny I}}(x)=|F_a| \langle \sum_i \vec{n}_i \delta(x-x_i+\xi(y_i)\rangle/L_y$, for a sharper view of the force distribution at the interface.
In  steady state, the mean active force has to be compensated by the mean repulsion between  particles (which may be used as a 
test for statistics in the ABP simulations, see Supplementary Info). 
Since 
%With the 
interfaces are along the $Y$ direction, only the $X$ component of the mean active force
is not null: we refer to it in all the following as $f^{\hbox{\tiny a}}_{\hbox{\tiny I}}(x)$.
%$\vec{f}^{V}_{\hbox{\tiny I}}(x)=- \langle \sum_{i,j} \vec{\nabla} V(r_{ij}) \delta(x-x_i+\xi(y_i)\rangle\L_y$, to get a sharp view of the forces at the interface, without the IS fluctuations that give the smoother mean averages $\vec{f}^{a}(x)$ and $\vec{f}_{V}(x)$.

The intrinsic profiles may be computed according to 
%in those of 
particles with a given coordination $\nu$, i.e. the number of neighbors  within  repulsion distance ($r_{ij}\leq 1.12 \sigma$). Most of the particles in the dense slab have $\nu=6$, with triangular lattice coordination, measured by the local hexatic order parameter, $\eta_j=\sum_{k} exp(6 i \theta_k)/6$, where $\theta_k$ is the orientation of the relative position $\vec{r}_{jk}$ for the ($k=1$ to $6$)  neighbors of particle $j$. For each particle in the dense phase, this parameter has a modulus $|\eta_j|\approx 1$, but its complex phase
reflects the local orientation of the neighbour. 
%star. 
The global  $\eta=\langle \sum_i \eta_i\rangle/N$, averaged over $N\gg 1$ particles, may have a modulus well below $1$, because of the different phase of each $\eta_i$. The hexatic 2D phase appears when $|\eta|$ is well above the noise ($\sim N^{-1/2}$).
In our ABP simulations we have calculated 
$\eta$ over the central half of the dense slab and, as it had been described before
~\cite{DiGregorio_PRL_2018,Caporusso_PRL_2020,digregorio_arXiv_2021,Paliwal_PhysRevResearch_2020}, near the borders there are large grains, with different phases of their $\eta$ averages. The global hexatic order is gradually lost by accumulation of $\nu=5$ and $7$ defects in a mesh of disclination lines denser at the edges.

Besides the intrinsic profiles, the CWT analysis gives information on the IS fluctuations.  As for equilibrium fluid interfaces \cite{Tarazona_CW_2007_PRL}, we get independent Gaussian distributions for the Fourier components $\hat{\xi}_q$ (S.I.) and the mean square amplitudes, written as $\langle |\hat{\xi}_q|^2\rangle= (q^2 \beta\gamma(q) L_y)^{-1}$, define a wavevector dependent surface (or line) tension $\gamma(q)$. In equilibrium systems these mean square fluctuations are proportional to $kT=\beta^{-1}$, and $\gamma(q)$ gives the macroscopic surface tension as the limit $\gamma_o=\gamma(0)$. In ABP interfaces we have to interpret $\beta \gamma(q)$ together, as an (inverse) measure of the IS fluctuations.
%, but only through an independent estimation of an {\sl effective temperature} could we get the (effective) $\gamma(q)$.{\bf ¿Hace falta esto ultimo? ¿o lo quitamos}

\section{\label{sec:Result}Results and Discussion}

\begin{figure}[h!]
\begin{center}
a)\includegraphics[width=3cm]{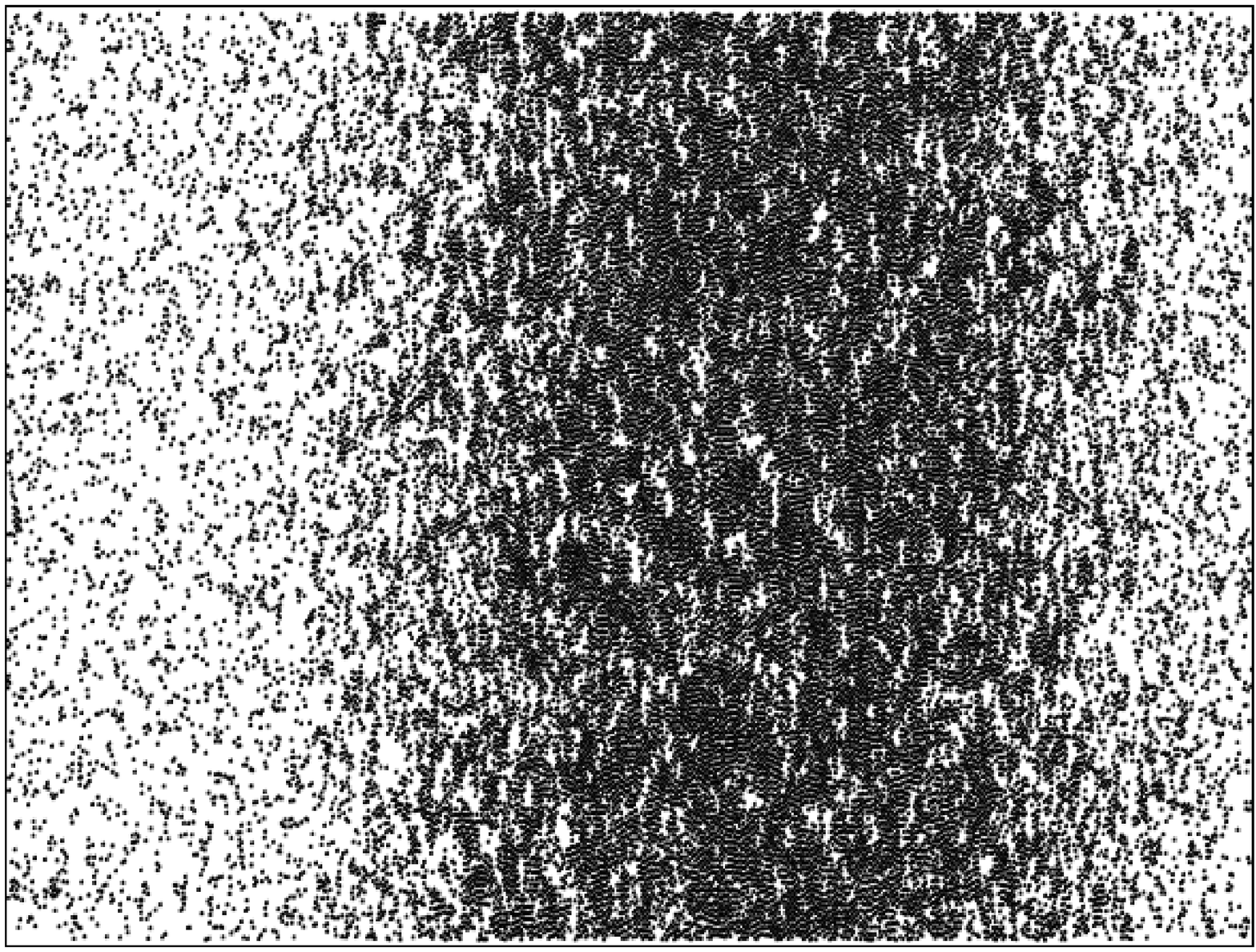}
b)\includegraphics[width=3cm]{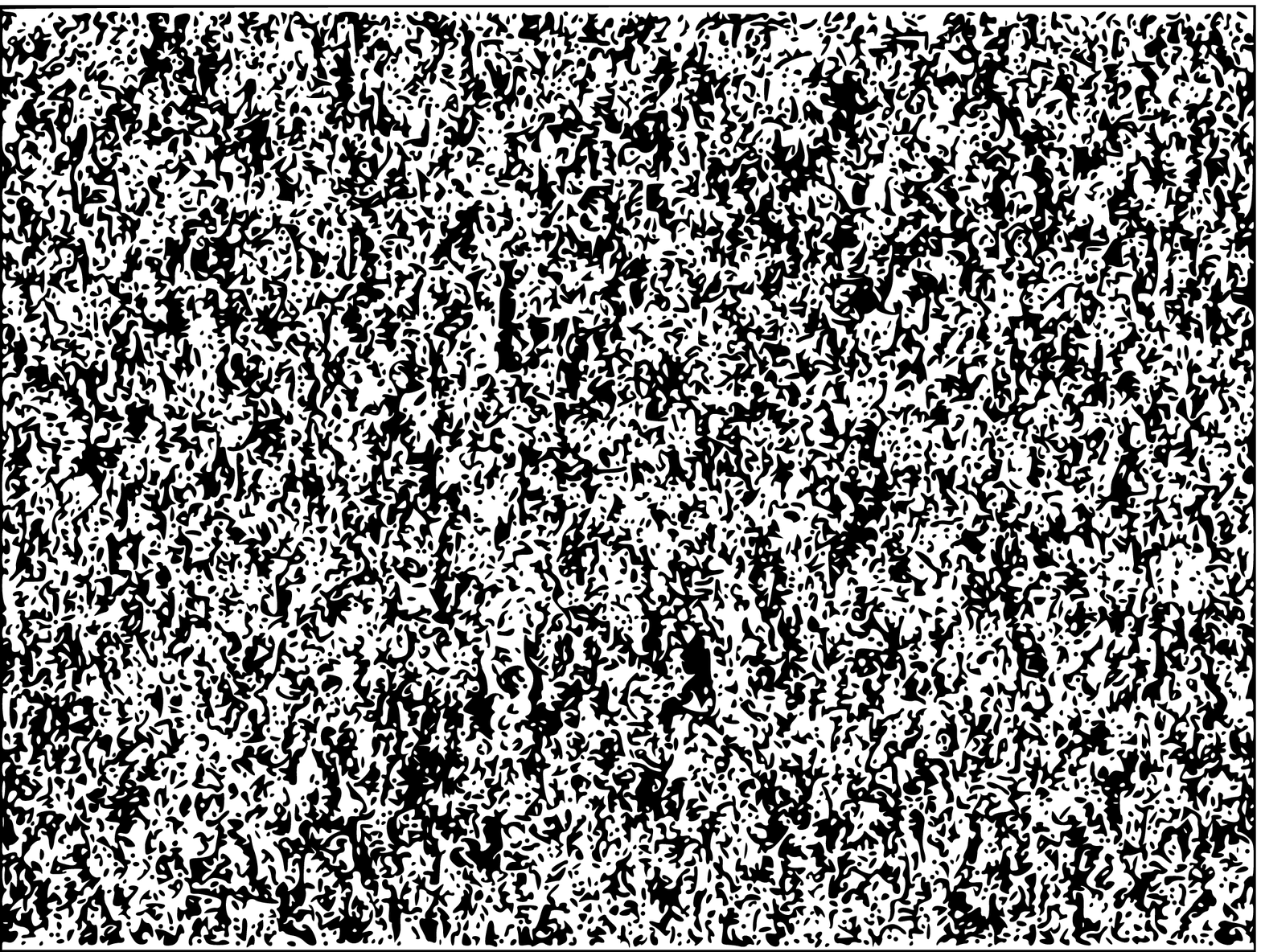}
c)\includegraphics[width=3cm]{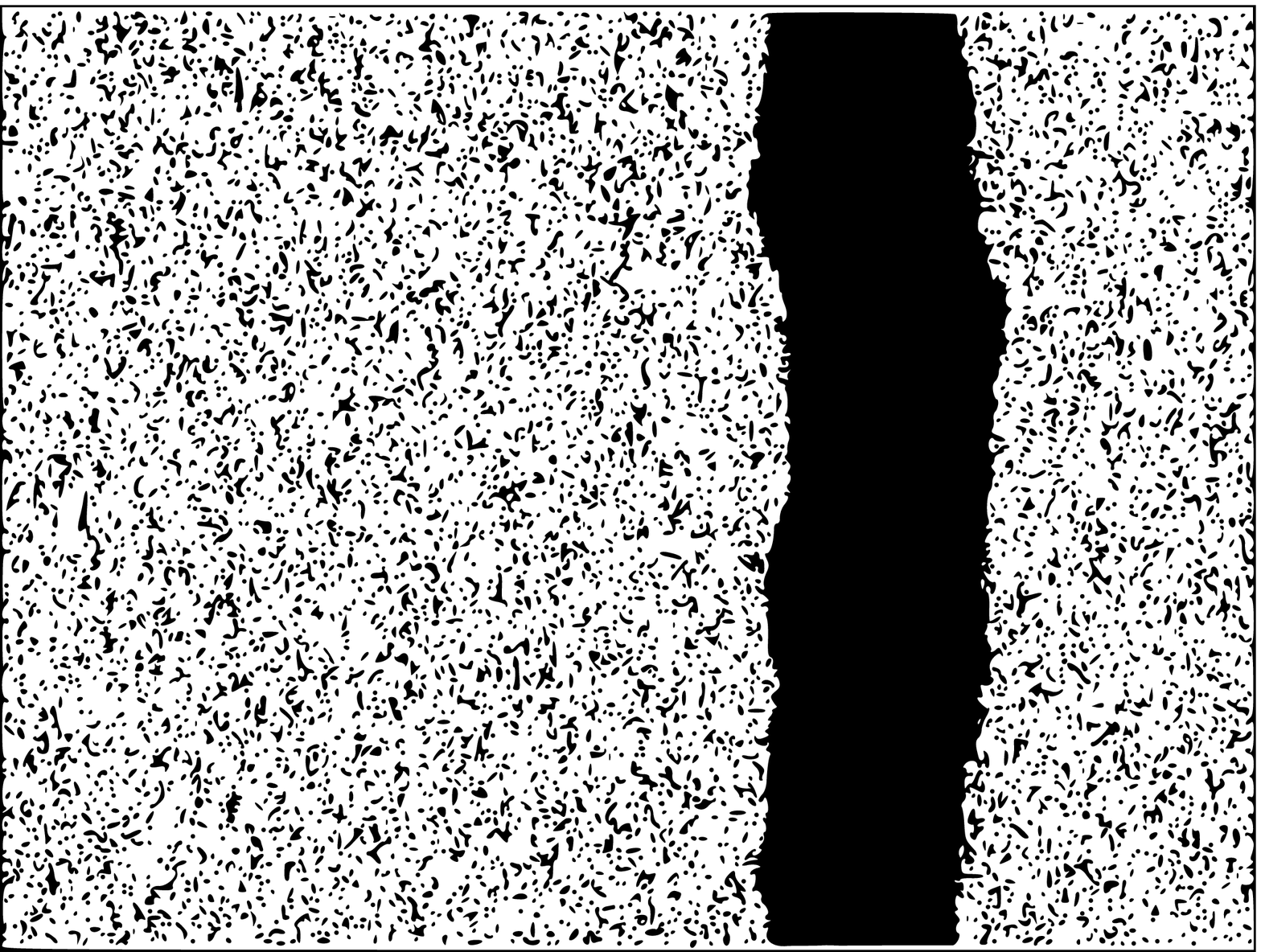}
%\caption{Snapshots of the last frame of simulations, using attractive ABPs with $F_a=\lbrace 0.1,1,3 \rbrace$ respectively.}
\caption{Snapshots 
%of the last frame of simulations, using 
of an attractive LJ-ABP suspension with $F_a=\lbrace 3, 30, 80 \rbrace$ respectively a,b,c.}
\label{fig:Small_Fa}
\end{center}
\end{figure}

Besides having prepared the initial configuration for a suspension of repulsive (WCA) ABPs, we study the effect of activity on a phase separated system of attractive (LJ) active particles. 
We prepare a phase separated system of  attractive particles at thermodynamic conditions at which its passive counterpart would phase separate. In Figure \ref{fig:Small_Fa} panel a) we observe that  adding a small activity to the attractive particles is not enough to break the dense-dilute  phase separation (also detected for the equilibrium passive system): for small values of activity  ($F_a < 3$) the dense phase  is stable as a band.

Similarly to previous studies \cite{Redner_aabp}, we   observe a reentrant phase behavior of the phase separated structure. When increasing the propulsion strength, 
%If  particles self-propel at medium speed i.e. 
$3<F_a<50$, (Figure \ref{fig:Small_Fa} panel a to b), activity overtakes attraction and  the band melts, leading to a gas phase.
%\begin{figure}[h!]
%\begin{center}
%\includegraphics[width=3.75cm]{FIGURES/Fa6_0t_100000000.png} \includegraphics[width=3.75cm]{FIGURES/Fa10_0t_073000000.png} \includegraphics[width=3.75cm]{FIGURES/Fa30_0t_083000000.png}
%\caption{Snapshots of the last frame of simulations, using attractive ABPs with $F_a=\lbrace 6,10,30 \rbrace$ respectively.}
%\label{fig:Medium_Fa}
%\end{center}
%\end{figure}
Strikingly, if we increase propulsion strength 
%the self-propel velocity 
even further ($F_a>40$), phase coexistence emerges again (Figure \ref{fig:Small_Fa} panel c). The system forms a more compact band, even though the mean density in the box is quite low $\rho=0.38$.
In what follows, we will consider either repulsively or attractively interacting particles at relatively high activity, where attraction is less relevant than propulsion. 

\begin{figure}[h!]
\centering
\includegraphics[width=0.8\linewidth]{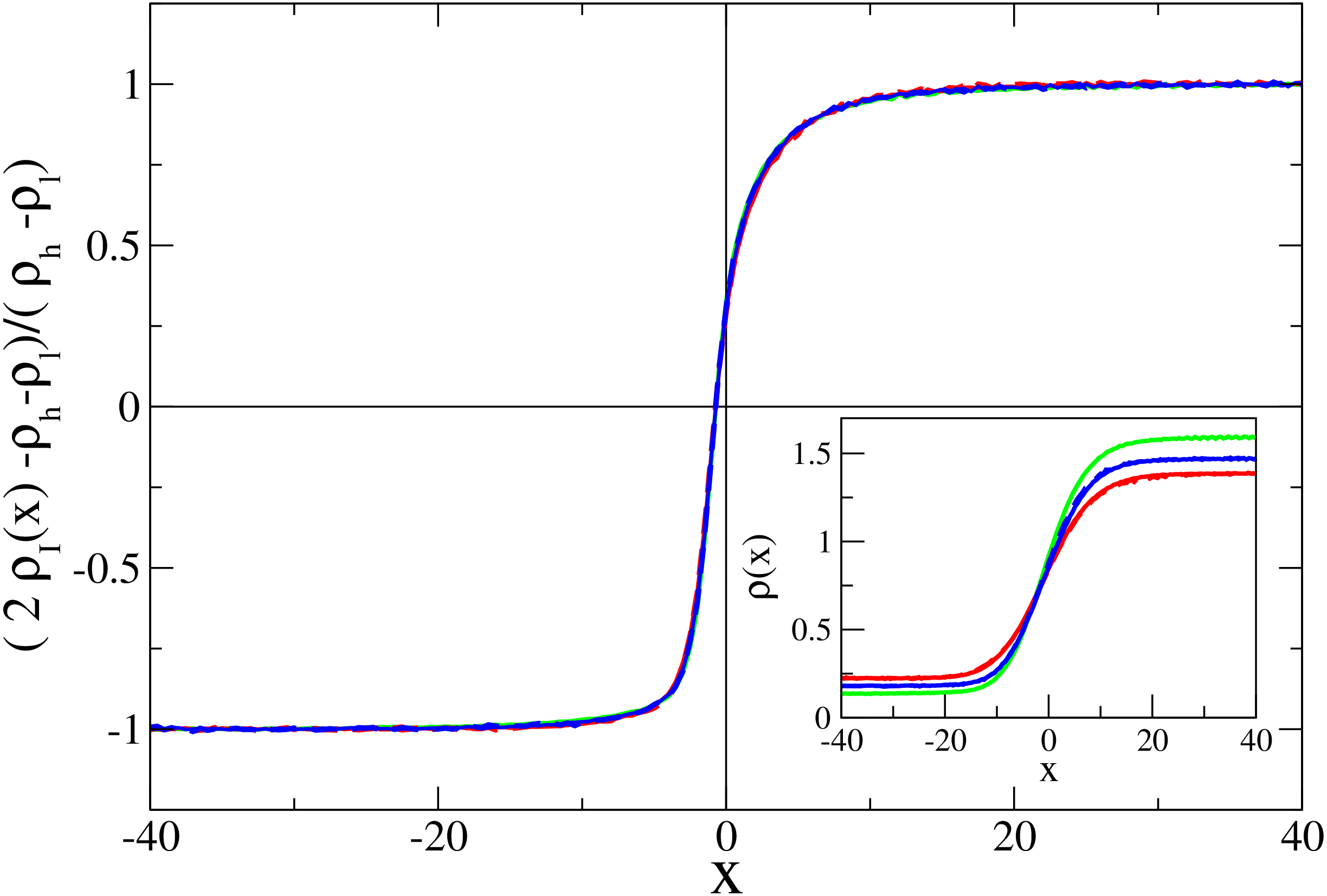}
\caption{(Mean (inset) and intrinsic (main panel)  density profiles for $F_a=60$ (red), $75$ (blue) and $120$ (green).
The IDP are scaled and shifted to get $\pm 1$ as the limit
values at the two sides of the interface, and the results
for all the $F_a$ values become nearly identical. Although indistinguishable, the full lines are the LJ-ABP results and the dashed line the WCA-ABP.
Units of the LJ $\sigma$ diameter are used. }
\label{fig:Profiles_Fps}
\end{figure}

Having prepared the initial configuration, we estimate the mean density profiles and the IDP for a dense-dilute MIPS formed by both repulsive and attractive ABP. 
Figure\ref{fig:Profiles_Fps} presents the IDPs for  three
values of  activity.  To eliminate the $F_a$ dependence
of the bulk densities, we  scale it and shift it  $(2\rho_{\hbox{\tiny I}}(x)-\rho_h-\rho_l)/\Delta\rho_{hl}$. 

All  scaled IPDs become nearly identical and  are asymmetric, with a  smoother decay towards the dense side. 
The inset in Fig.\ref{fig:Profiles_Fps} shows the mean density profiles that become broader as we reduce $F_a$.
The mean density profiles $\rho(x)$ become smoother for  decreasing activity because of the larger IS fluctuations, rather than from any local change in the interfacial region, other than the scaling of the bulk densities. 
Interestingly, the results obtained for repulsive particles coincide with the ones for attractive ones. For the chosen activity range, the inter-particle interaction plays no role in the mean density profile.

\begin{figure}[h!]
\centering
\includegraphics[width=\linewidth]{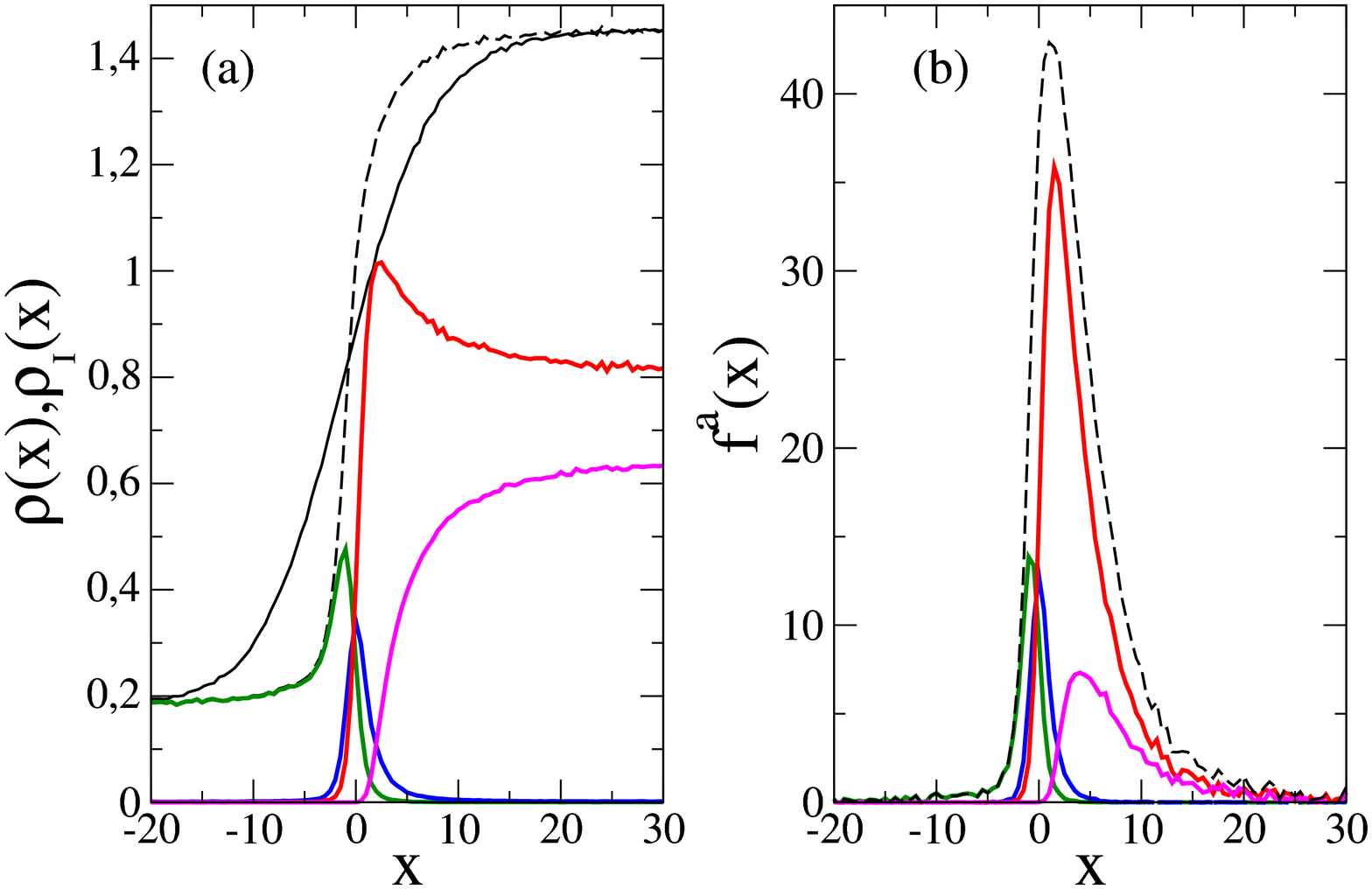}
\caption{Density (panel a) and active force (panel b) profiles for the LJ-ABP with $F_a=80$. Full black line (a) mean density profiles. The intrinsic profiles (dashed black lines in both panel) are
split in particles with $\nu\leq 3$ (green), $\nu=4$ (blue),
$\nu=5$ or $7$ (red), and $\nu=6$ (magenta) coordination numbers. 
}
\label{fig:Profiles_mx}
\end{figure}

To better characterise the interface, we study the density and the active force profiles  for different coordination numbers, when the dense-dilute MIPS is made of either repulsive or attractive particles.
Fig.\ref{fig:Profiles_mx}a represents mean density profiles $\rho(x)$ (continuous lines) and intrinsic density profiles $\rho_I(x)$ (dashed lines) for coordination numbers 
$\nu \le 3$ (green lines), $\nu =4$ (blue lines), $\nu =5$ or $\nu =7$  (red lines) and 
$\nu =6$ (magenta lines).

In the low density region, comparing the coloured lines, we conclude that the low density phase is mostly made of particles with $\nu\leq 3$ (green line). Moreover,  the rapid increase of $\rho_{\hbox{\tiny I}}(x)$ (dashed line) from the $\rho_l$ bulk, comes from the accumulation of $\nu=2-3$ particles at the edge of the dense slab. Particles with $\nu=4$ (blue line) are distributed as a very narrow peak at  $x=\xi(y)$, which provides an independent test for our IS definition. In the high density phase we get $\nu\geq 5$ (red and magenta lines) and the very slow raise of $\rho_{\hbox{\tiny I}}(x)$ is produced mainly by the gradual decay of five-fold correlated particles, as we move into the bulk.

Fig.\ref{fig:Profiles_mx}b) represents the active force profiles $f_a(x)$ for the mean density  (continuous lines) and for the intrinsic density  (dashed lines), and for coordination numbers 
$\nu \le 3$ (green lines), $\nu =4$ (blue lines), $\nu =5$ or $\nu =7$  (red lines) and 
$\nu =6$ (magenta lines). 
When $\nu \le 4$, the profile is peaked at the low density region,  
whereas when $\nu \le 5$ the peak becomes much higher and moves towards the high density region. 
The high density slab is compressed to a higher pressure than the lower density phase, which make the MIPS interface to be mechanically very different from an equilibrium liquid-vapor surface. Notice this interpretation in terms of different pressure between coexisting phases is that of Ref.~\cite{Omar_PRE_2020},, while other authors used to include the effect of the active force as part of the stress tensor, rather than as an effective external force, and get equal pressure between coexisting phases. 
Both approaches are equally valid, if treated consistently.

The active force profiles demonstrate how the correlation with the particle position  rectifies the random orientation of the active force and creates  the effect of a mean external potential at the interface ~\cite{Omar_PRE_2020}. 
The rectification of the active force was interpreted as the result of the swimming pressure of the low density phase on the edge of the dense slab. When the active force on these highly mobile particles points inward ($n_x>0$) it is arrested by the repulsion of the particles in the slab, but when $n_x<0$ the particles are free to move out of the interface, they dissipate the active force as friction, and become scarcer at the interface.

 Our intrinsic view for the active force profiles allows a much sharper characterization and gives a different perspective. 
The rectification of the active force comes mainly from the denser side of the interface, rather than from the external layer of freely moving particles.
Given that mobility in the dense phase is small, the local active force’s rectification happens on time scales longer than the typical characteristic times of the angular diffusion.  
The $x$ component of the active force has an asymmetric peak %of $f^{a}_{\hbox{\tiny I}}(x)$ is 
extended over distances $-5\sigma \leq x \leq 20 \sigma$ 
from the IS, similar to the asymmetry of the intrinsic density $\rho_{\hbox{\tiny I}}(x)$. The splitting
of $f^{\hbox{\tiny a}}_{\hbox{\tiny I}}(x)$ in terms of $\nu$ shows that the rectification of the random active force comes mainly from particles with crystalline correlation ($\nu=6$) or at lattice point defects ($\nu=5$, $7$).   Apparently, these particles should be too tightly caged by their neighbors to get the asymmetric mobility that rectifies their active force as a swimming pressure.

The explanation for this feature comes from the slow raise of $\rho_{I}(x)$ towards the dense phase, which signals the accumulation of point defects (mostly particles with $\nu=5$ coordination) over that thick region $\sim 20 \sigma$ at the edge of the slab. Such accumulation of lattice defects implies the presence of disclination lines, or hexatic grain boundaries, that gradually break the nearly perfect 2D crystal correlation in the interior of the dense slab when we approach the interface \cite{Redner_PRL_2013,Kamser_NatCom_2018,DiGregorio_PRL_2018,digregorio_arXiv_2021,Paliwal_PhysRevResearch_2020,Caporusso_PRL_2020}. 
The mobility of  particles along these grain boundaries should be asymmetric, 
%been
more easily (collectively) shifted by the active force when $\vec{n}_i$ points towards the interface than when it pushes towards the interior of the slab.

The gradual increase of the compression, as we move towards the center of the slab, reduces the defects, slowly increasing the density $\rho_{\hbox{\tiny I}}(x)$, and that reduces the rectification of the active force $F_a \vec{n}$. At the inner part of the slab the
distribution of lattice defects and dislocation lines becomes statistically homogeneous, the rectification of the active force disappears, and (consistently) the central part of the dense slab becomes a homogeneous hexatic phase characterised by  bulk density.

\begin{figure}[h!]
\centering
\includegraphics[width=\linewidth]{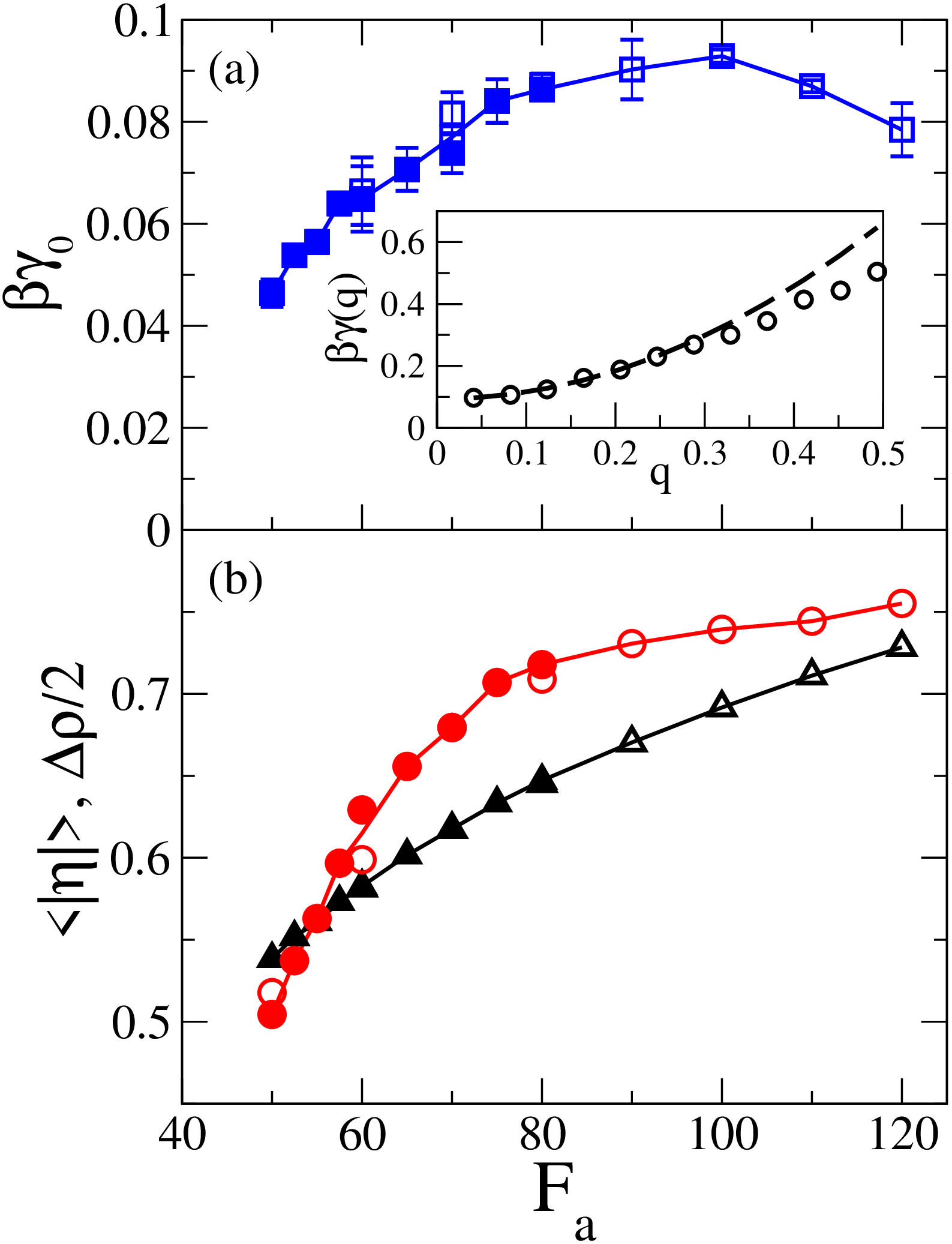}
\caption{ (a) The value of $\beta\gamma_o$ extracted from the mean fluctuations at the two edges of the slab (error bars give the difference between the two sides). (Inset) The function $\beta \gamma(q)$, for the wavevectors $q$ in our simulation cell,
for $F_a=100$, with its quadratic fit to $\beta \gamma_o+ \beta K q^2$, to extract $\beta \gamma_0$  represented in the blue symbols. 
(b) 
Half the density gap, $(\rho_{h}-\rho_l)/2$ (black triangles),and the mean modulus of the hexatic order parameter $\langle|\eta|\rangle$ calculated over the central part of the dense slab (red circle), as functions of the active force $F_a$.
The solids symbols are the LJ-ABP results and the empty symbols the WCA-ABP, the lines are eyes guide.
}
\label{fig:Interfacial_Tension}
\end{figure}

 Finally, within the CWT analysis we compute the mean square amplitude of the IS fluctuations, through the 
function $\beta \gamma(q)=(q^2 \langle |\hat{\xi}_q|^2\rangle L_y)^{-1}$, as shown in Fig.~\ref{fig:Interfacial_Tension} (top panel). 

To study the limit of $\beta \gamma(q)$ at low $q$, we use the following fit
$\beta \gamma(q)\approx \beta\gamma_o + \beta\kappa q^2$, that is quite accurate and gives robust results (Fig.~\ref{fig:Interfacial_Tension} inset).
%The fit $\beta \gamma(q)\approx \beta\gamma_o + \beta\kappa q^2$ is quite accurate (Fig.~\ref{fig:Interfacial_Tension} inset panel) and it gives a robust  result for the limit $\beta \gamma_o$ of $\beta \gamma(q)$ at low $q$. 
Figure~\ref{fig:Interfacial_Tension} - top panel presents $\beta \gamma_o$ as a function of $F_a$. The error bars have been computed 
to give the difference between the results obtained on the two edges of the slab. 

In a thermal equilibrium liquid-vapor interface, the $\beta$ factor may be removed to get (from the analysis of the surface fluctuations) the same $\gamma_o$ that gives the mechanical work (the free energy change) for increasing the area. 
In an ABP MIPS, the mechanical evaluation of $\gamma_o$ has led to surprisingly negative values\cite{PRLSpeck}. 
The interpretation of the MIPS as coexistence
of two phases with a bulk pressure difference, self-maintained by the rectification of the active force at the interfaces~\cite{Omar_PRE_2020}, voids that
controversy. 
Even if we could get rid of the $\beta$ factor, our $\gamma_o$ could not be used to get the mechanical work of a global deformation of the system without including
the pressure difference between the two phases.  
Our  capillary surface tension, $\beta \gamma_o$, is always positive, and it would vanish only if the interfacial fluctuations become unbounded and destroy the MIPS.

To conclude, we study how  MIPS disappears for $F_a \leq 50$. The usual description is presented in terms of the density gap $\Delta \rho_{hl}$ in the $(\rho,F_a)$ phase diagrams~\cite{DiGregorio_PRL_2018}, as for equilibrium phase transitions.  Our analysis adds two clearer clues with the rapid decay of the hexatic order parameter $\langle|\eta|\rangle$ over the bulk of the dense phase, and of $\beta \gamma_o$ at the interfaces. Figure~\ref{fig:Interfacial_Tension}- bottom panel reports  half the density gap, $(\rho_{h}-\rho_l)/2$ and the mean modulus of the hexatic order parameter $\langle|\eta|\rangle$ calculated over the central part of the dense slab, as functions of the active force $F_a$ ($50\leq F_a\leq 120$). The dependence of  $\Delta \rho_{hl}$ over that range is very smooth, without any signal that MIPS may be close to its end for $F_a\lesssim 50$. However, both $\langle|\eta|\rangle$ and $\beta \gamma_o$ show a clear slope change at $F_a\approx 80$ and  decay rapidly for lower activity. 

 The connection between these two magnitudes may be rationalized when we interpret the MIPS of the ABP in the terms suggested by the
intrinsic density and force profiles. A weaker active force reduces the pressure on the dense slab, and its density decreases mainly by changing the mean nearest neighbor distance. When that effect is scaled out, as in Fig.~\ref{fig:Profiles_Fps} we get very similar IDP for all the $F_a$. However, the hexatic order at the inner part of the slab is rapidly weakened as the density decreases. Large local fluctuations in the hexatic order at the center of the slab would go against the rectification of the random active force, since the disclination lines could reach the inner part of the slab, instead of being mainly restricted to the grain boundaries over the width $\sim 20\sigma$ at the edges of the 
slab where $f^{\hbox{\tiny a}}(x)$ has its main contribution. %In turn, that reduction of the effective potential from the active force would weaken still further the compression of the slab, and its hexatic order. The self-maintained pressure gradient across the interfaces, based on the concentration of crystal grain boundaries at the edges of the slab, may fall down as the end of the MIPS.
The intuitive image that appears from our analysis is
that when the active force, rectified through the asymmetric mobility of the particles along the grain boundaries, becomes too weak to keep the inner hexatic order of the dense slab, the grain boundaries may percolate from side to side of the slab and act as wedges that break the dense phase  in pieces, to end with the MIPS.

\section{Conclusions}

Most of the theoretical concepts and tools we have implemented in our work  have been recently used  by different authors. Patch et al.\cite{Patch_SoftMat_2018} presented an intrinsic surface analysis of the ABP interface similar to ours; the full phase diagram, including the analysis of the hexatic order, had been studied by several groups \cite{Redner_PRL_2013,Kamser_NatCom_2018,DiGregorio_PRL_2018,digregorio_arXiv_2021,Paliwal_PhysRevResearch_2020,Caporusso_PRL_2020}, and the interpretation of  MIPS as an effect of an effective external potential, created by the active force across density gradient, was proposed by Omar et al.\cite{Omar_PRE_2020}. Our contribution relies on  unifying  all these concepts, together with adding a novel view of the mechanism for MIPS coming from our CWT analysis.

The force generated at the interface, as a rectification of the random direction of active force, comes mainly from the dense side of the interface, through a rather thick layer (of about $20$ particle diameters). This is shown in the intrinsic
force and density profiles. The latter with an amazing similarity in their scaled shapes over the whole range of MIPS. 
The image of the swimming pressure, created by nearly free particles in the low density phase under the effects of the random active force, breaks down for the tightly caged particles that contribute most to the compression of the dense phase slab.
The rectification of the random active force, to generate an effective external potential, has to come from collective rearrangements of the caged particles, at the hexatic phase grains which had been recently observed and characterized \cite{DiGregorio_PRL_2018, digregorio_arXiv_2021,Paliwal_PhysRevResearch_2020,Caporusso_PRL_2020} as an important feature of the MIPS in 2D-ABP. There is clear correlation between the rapid loss of the hexatic order at the inner part of that slab; the decrease of $\beta \gamma(0)$ (i.e. the growth of the interfacial fluctuations), and the end of the MIPS at $F_a \lesssim 50$.

As far as we may ascertain, the end of the MIPS comes for $F_a$ just below $50$, when the compression by the active force cannot keep the hexatic order at the inner of the slab. That is correlated with the decay of $\beta \gamma_o$, i.e. the
growth of the geometrical fluctuations of the interface. In their dependence with active force, both the hexatic order parameter $\langle|\eta|\rangle$ and  $\beta \gamma_o$ show a change at $F_a\approx 80$, that is associated to the pinning of the hexatic phase by the direction of the boundaries. Although with  different interactions and parameters, this agrees qualitatively with the distinction between "hexatic" and "solid"
for the structure of the dense coexisting phase in the full phase diagram~\cite{DiGregorio_PRL_2018}.
Nevertheless, the accurate characterization of that region of the ABP phase diagram faces  great difficulties, because  large spatial and temporal scales are needed to simulate the 2D hexatic crystal slab, with a thick premelted film at each border, and approaching its bulk melting conditions. In this respect, the recent work by Caporusso et al.~\cite{Caporusso_PRL_2020} for the nucleation and growth of the hexatic grains in large ABP simulations indicates that the macroscopic separation of an inner homogeneous dense phase may be strongly propitiated, and perhaps created, by its percolation as a slab, through the periodic boundary conditions.

Finally, at the high $F_a$ values explored here, our use of the attractive interaction between  particles coincide with results obtained for repulsive particles. However, at much lower $F_a$, the  system of attractive particles shows a liquid-vapor coexistence,
that smoothly goes  to the  thermal equilibrium transition at $F_a=0$. The application of CWT-based analysis methods  to those weakly-active LJ systems may be as useful as it has been in the strongly-active case studied here.

%\section*{Author Contributions}
%{\bf escribirlo com querais espeficando tareas o quitarlo}
%We strongly encourage authors to include author contributions and recommend using \href{https://casrai.org/credit/}{CRediT} for standardised contribution descriptions. Please refer to our general \href{https://www.rsc.org/journals-books-databases/journal-authors-reviewers/author-responsibilities/}{author guidelines} for more information about authorship.

\section*{Conflicts of interest}
There are no conflicts to declare.

\section*{Acknowledgements}
We acknowledge the support of the Spanish
Secretariat for Research, Development and Innovation
(Grants No. FIS2017-86007-C3, PID2020-117080RB-C52,   FPU2015/0248, and PID2019-105343GB-I00) and
from the Maria de Maeztu Programme for Units of Excellence in R\&D
(CEX2018-000805-M).
%{\bf Pedro cual es la referencia del nuevo Programa Ramiro Maeztu}

%%%END OF MAIN TEXT%%%

%The \balance command can be used to balance the columns on the final page if desired. It should be placed anywhere within the first column of the last page.
%\balance

%If notes are included in your references you can change the title from 'References' to 'Notes and references' using the following command:
%\renewcommand\refname{Notes and references}

%%%REFERENCES%%%

\clearpage
%\bibliography{ABP2021}

\begin{thebibliography}{59}%
\makeatletter
\providecommand \@ifxundefined [1]{%
 \@ifx{#1\undefined}
}%
\providecommand \@ifnum [1]{%
 \ifnum #1\expandafter \@firstoftwo
 \else \expandafter \@secondoftwo
 \fi
}%
\providecommand \@ifx [1]{%
 \ifx #1\expandafter \@firstoftwo
 \else \expandafter \@secondoftwo
 \fi
}%
\providecommand \natexlab [1]{#1}%
\providecommand \enquote  [1]{``#1''}%
\providecommand \bibnamefont  [1]{#1}%
\providecommand \bibfnamefont [1]{#1}%
\providecommand \citenamefont [1]{#1}%
\providecommand \href@noop [0]{\@secondoftwo}%
\providecommand \href [0]{\begingroup \@sanitize@url \@href}%
\providecommand \@href[1]{\@@startlink{#1}\@@href}%
\providecommand \@@href[1]{\endgroup#1\@@endlink}%
\providecommand \@sanitize@url [0]{\catcode `\\12\catcode `\$12\catcode
  `\&12\catcode `\#12\catcode `\^12\catcode `\_12\catcode `\%12\relax}%
\providecommand \@@startlink[1]{}%
\providecommand \@@endlink[0]{}%
\providecommand \url  [0]{\begingroup\@sanitize@url \@url }%
\providecommand \@url [1]{\endgroup\@href {#1}{\urlprefix }}%
\providecommand \urlprefix  [0]{URL }%
\providecommand \Eprint [0]{\href }%
\providecommand \doibase [0]{http://dx.doi.org/}%
\providecommand \selectlanguage [0]{\@gobble}%
\providecommand \bibinfo  [0]{\@secondoftwo}%
\providecommand \bibfield  [0]{\@secondoftwo}%
\providecommand \translation [1]{[#1]}%
\providecommand \BibitemOpen [0]{}%
\providecommand \bibitemStop [0]{}%
\providecommand \bibitemNoStop [0]{.\EOS\space}%
\providecommand \EOS [0]{\spacefactor3000\relax}%
\providecommand \BibitemShut  [1]{\csname bibitem#1\endcsname}%
\let\auto@bib@innerbib\@empty
%</preamble>
\bibitem [{\citenamefont {Stenhammar}\ \emph {et~al.}(2014)\citenamefont
  {Stenhammar}, \citenamefont {Marenduzzo}, \citenamefont {Allen},\ and\
  \citenamefont {M.E.}}]{Stenhammar2014}%
  \BibitemOpen
  \bibfield  {author} {\bibinfo {author} {\bibfnamefont {J.}~\bibnamefont
  {Stenhammar}}, \bibinfo {author} {\bibfnamefont {D.}~\bibnamefont
  {Marenduzzo}}, \bibinfo {author} {\bibfnamefont {R.}~\bibnamefont {Allen}}, \
  and\ \bibinfo {author} {\bibfnamefont {C.}~\bibnamefont {M.E.}},\ }\href@noop
  {} {\bibfield  {journal} {\bibinfo  {journal} {Soft Matter}\ }\textbf
  {\bibinfo {volume} {10}},\ \bibinfo {pages} {1489} (\bibinfo {year}
  {2014})}\BibitemShut {NoStop}%
\bibitem [{\citenamefont {Julian~Bialke}\ \emph {et~al.}(2015)\citenamefont
  {Julian~Bialke}, \citenamefont {Siebert}, \citenamefont {Lowen},\ and\
  \citenamefont {Speck}}]{PRLSpeck}%
  \BibitemOpen
  \bibfield  {author} {\bibinfo {author} {\bibfnamefont {J.}~\bibnamefont
  {Julian~Bialke}}, \bibinfo {author} {\bibfnamefont {J.~T.}\ \bibnamefont
  {Siebert}}, \bibinfo {author} {\bibfnamefont {H.}~\bibnamefont {Lowen}}, \
  and\ \bibinfo {author} {\bibfnamefont {T.}~\bibnamefont {Speck}},\
  }\href@noop {} {\bibfield  {journal} {\bibinfo  {journal} {Phys. Rev. lett.}\
  }\textbf {\bibinfo {volume} {115}},\ \bibinfo {pages} {098301} (\bibinfo
  {year} {2015})}\BibitemShut {NoStop}%
\bibitem [{\citenamefont {Omar}\ \emph {et~al.}(2020)\citenamefont {Omar},
  \citenamefont {Wang},\ and\ \citenamefont {Brady}}]{Omar_PRE_2020}%
  \BibitemOpen
  \bibfield  {author} {\bibinfo {author} {\bibfnamefont {A.~K.}\ \bibnamefont
  {Omar}}, \bibinfo {author} {\bibfnamefont {Z.-G.}\ \bibnamefont {Wang}}, \
  and\ \bibinfo {author} {\bibfnamefont {J.~F.}\ \bibnamefont {Brady}},\
  }\href@noop {} {\bibfield  {journal} {\bibinfo  {journal} {Phys. Rev. E}\
  }\textbf {\bibinfo {volume} {101}},\ \bibinfo {pages} {012604} (\bibinfo
  {year} {2020})}\BibitemShut {NoStop}%
\bibitem [{\citenamefont {Bechinger}\ \emph {et~al.}(2016)\citenamefont
  {Bechinger}, \citenamefont {Di~Leonardo}, \citenamefont {Lowen},
  \citenamefont {Reichhardt}, \citenamefont {Volpe},\ and\ \citenamefont
  {Volpe}}]{Bechinger2016}%
  \BibitemOpen
  \bibfield  {author} {\bibinfo {author} {\bibfnamefont {C.}~\bibnamefont
  {Bechinger}}, \bibinfo {author} {\bibfnamefont {R.}~\bibnamefont
  {Di~Leonardo}}, \bibinfo {author} {\bibfnamefont {H.}~\bibnamefont {Lowen}},
  \bibinfo {author} {\bibfnamefont {C.}~\bibnamefont {Reichhardt}}, \bibinfo
  {author} {\bibfnamefont {G.}~\bibnamefont {Volpe}}, \ and\ \bibinfo {author}
  {\bibfnamefont {G.}~\bibnamefont {Volpe}},\ }\href@noop {} {\bibfield
  {journal} {\bibinfo  {journal} {Reviews of Modern Physics.}\ }\textbf
  {\bibinfo {volume} {88}},\ \bibinfo {pages} {045006} (\bibinfo {year}
  {2016})}\BibitemShut {NoStop}%
\bibitem [{\citenamefont {Redner}\ \emph
  {et~al.}(2013{\natexlab{a}})\citenamefont {Redner}, \citenamefont {Hagan}, ,\
  and\ \citenamefont {Baskaran}}]{RednerPRL}%
  \BibitemOpen
  \bibfield  {author} {\bibinfo {author} {\bibfnamefont {G.~S.}\ \bibnamefont
  {Redner}}, \bibinfo {author} {\bibfnamefont {M.~F.}\ \bibnamefont {Hagan}}, ,
  \ and\ \bibinfo {author} {\bibfnamefont {A.}~\bibnamefont {Baskaran}},\
  }\href@noop {} {\bibfield  {journal} {\bibinfo  {journal} {Physical Review
  Letters}\ }\textbf {\bibinfo {volume} {110}},\ \bibinfo {pages} {055701}
  (\bibinfo {year} {2013}{\natexlab{a}})}\BibitemShut {NoStop}%
\bibitem [{\citenamefont {Redner}\ \emph
  {et~al.}(2013{\natexlab{b}})\citenamefont {Redner}, \citenamefont
  {Baskaran},\ and\ \citenamefont {Hagan}}]{Redner_aabp}%
  \BibitemOpen
  \bibfield  {author} {\bibinfo {author} {\bibfnamefont {G.~S.}\ \bibnamefont
  {Redner}}, \bibinfo {author} {\bibfnamefont {A.}~\bibnamefont {Baskaran}}, \
  and\ \bibinfo {author} {\bibfnamefont {M.~F.}\ \bibnamefont {Hagan}},\ }\href
  {\doibase 10.1103/PhysRevE.88.012305} {\bibfield  {journal} {\bibinfo
  {journal} {Phys. Rev. E}\ }\textbf {\bibinfo {volume} {88}},\ \bibinfo
  {pages} {012305} (\bibinfo {year} {2013}{\natexlab{b}})}\BibitemShut
  {NoStop}%
\bibitem [{\citenamefont {Mognetti}\ \emph {et~al.}(2013)\citenamefont
  {Mognetti}, \citenamefont {Saric}, \citenamefont {Angioletti-Uberti},
  \citenamefont {Cacciuto}, \citenamefont {Valeriani},\ and\ \citenamefont
  {Frenkel}}]{Mognetti2013}%
  \BibitemOpen
  \bibfield  {author} {\bibinfo {author} {\bibfnamefont {B.~M.}\ \bibnamefont
  {Mognetti}}, \bibinfo {author} {\bibfnamefont {A.}~\bibnamefont {Saric}},
  \bibinfo {author} {\bibfnamefont {S.}~\bibnamefont {Angioletti-Uberti}},
  \bibinfo {author} {\bibfnamefont {A.}~\bibnamefont {Cacciuto}}, \bibinfo
  {author} {\bibfnamefont {C.}~\bibnamefont {Valeriani}}, \ and\ \bibinfo
  {author} {\bibfnamefont {D.}~\bibnamefont {Frenkel}},\ }\href@noop {}
  {\bibfield  {journal} {\bibinfo  {journal} {Physical Review Letters}\
  }\textbf {\bibinfo {volume} {111}},\ \bibinfo {pages} {245702} (\bibinfo
  {year} {2013})}\BibitemShut {NoStop}%
\bibitem [{\citenamefont {Alarcon}\ \emph {et~al.}(2017)\citenamefont
  {Alarcon}, \citenamefont {Valeriani},\ and\ \citenamefont
  {Pagonabarraga}}]{AlarconSoftmatter}%
  \BibitemOpen
  \bibfield  {author} {\bibinfo {author} {\bibfnamefont {F.}~\bibnamefont
  {Alarcon}}, \bibinfo {author} {\bibfnamefont {C.}~\bibnamefont {Valeriani}},
  \ and\ \bibinfo {author} {\bibfnamefont {I.}~\bibnamefont {Pagonabarraga}},\
  }\href@noop {} {\bibfield  {journal} {\bibinfo  {journal} {Soft matter}\
  }\textbf {\bibinfo {volume} {13}},\ \bibinfo {pages} {814} (\bibinfo {year}
  {2017})}\BibitemShut {NoStop}%
\bibitem [{\citenamefont {Sarkar}\ \emph {et~al.}(2021)\citenamefont {Sarkar},
  \citenamefont {Gompper},\ and\ \citenamefont {Elgeti}}]{Sarkar2021}%
  \BibitemOpen
  \bibfield  {author} {\bibinfo {author} {\bibfnamefont {D.}~\bibnamefont
  {Sarkar}}, \bibinfo {author} {\bibfnamefont {G.}~\bibnamefont {Gompper}}, \
  and\ \bibinfo {author} {\bibfnamefont {J.}~\bibnamefont {Elgeti}},\
  }\href@noop {} {\bibfield  {journal} {\bibinfo  {journal} {Commun. Phys.}\
  }\textbf {\bibinfo {volume} {4}},\ \bibinfo {pages} {36} (\bibinfo {year}
  {2021})}\BibitemShut {NoStop}%
\bibitem [{\citenamefont {Palacci}\ \emph {et~al.}(2013)\citenamefont
  {Palacci}, \citenamefont {Sacanna}, \citenamefont {Steinberg}, \citenamefont
  {Pine},\ and\ \citenamefont {Chaikin}}]{palacci}%
  \BibitemOpen
  \bibfield  {author} {\bibinfo {author} {\bibfnamefont {J.}~\bibnamefont
  {Palacci}}, \bibinfo {author} {\bibfnamefont {S.}~\bibnamefont {Sacanna}},
  \bibinfo {author} {\bibfnamefont {A.~P.}\ \bibnamefont {Steinberg}}, \bibinfo
  {author} {\bibfnamefont {D.~J.}\ \bibnamefont {Pine}}, \ and\ \bibinfo
  {author} {\bibfnamefont {P.~M.}\ \bibnamefont {Chaikin}},\ }\href@noop {}
  {\bibfield  {journal} {\bibinfo  {journal} {Science}\ }\textbf {\bibinfo
  {volume} {339}},\ \bibinfo {pages} {936} (\bibinfo {year}
  {2013})}\BibitemShut {NoStop}%
\bibitem [{\citenamefont {Buttinoni}\ \emph {et~al.}(2013)\citenamefont
  {Buttinoni}, \citenamefont {Bialke}, \citenamefont {Kummel}, \citenamefont
  {Lowen}, \citenamefont {Bechinger},\ and\ \citenamefont
  {Speck}}]{Buttinoni2013}%
  \BibitemOpen
  \bibfield  {author} {\bibinfo {author} {\bibfnamefont {I.}~\bibnamefont
  {Buttinoni}}, \bibinfo {author} {\bibfnamefont {J.}~\bibnamefont {Bialke}},
  \bibinfo {author} {\bibfnamefont {F.}~\bibnamefont {Kummel}}, \bibinfo
  {author} {\bibfnamefont {H.}~\bibnamefont {Lowen}}, \bibinfo {author}
  {\bibfnamefont {C.}~\bibnamefont {Bechinger}}, \ and\ \bibinfo {author}
  {\bibfnamefont {T.}~\bibnamefont {Speck}},\ }\href@noop {} {\bibfield
  {journal} {\bibinfo  {journal} {Physical review letters}\ }\textbf {\bibinfo
  {volume} {110}},\ \bibinfo {pages} {238301} (\bibinfo {year}
  {2013})}\BibitemShut {NoStop}%
\bibitem [{\citenamefont {Theurkauff}\ \emph {et~al.}(2012)\citenamefont
  {Theurkauff}, \citenamefont {Cottin-Bizonne}, \citenamefont {Palacci},
  \citenamefont {Ybert},\ and\ \citenamefont {Bocquet}}]{Theurkauff}%
  \BibitemOpen
  \bibfield  {author} {\bibinfo {author} {\bibfnamefont {I.}~\bibnamefont
  {Theurkauff}}, \bibinfo {author} {\bibfnamefont {C.}~\bibnamefont
  {Cottin-Bizonne}}, \bibinfo {author} {\bibfnamefont {J.}~\bibnamefont
  {Palacci}}, \bibinfo {author} {\bibfnamefont {C.}~\bibnamefont {Ybert}}, \
  and\ \bibinfo {author} {\bibfnamefont {L.}~\bibnamefont {Bocquet}},\
  }\href@noop {} {\bibfield  {journal} {\bibinfo  {journal} {Physical Review
  Letters}\ }\textbf {\bibinfo {volume} {108}},\ \bibinfo {pages} {268303}
  (\bibinfo {year} {2012})}\BibitemShut {NoStop}%
\bibitem [{\citenamefont {S.~Hermann}\ \emph {et~al.}(2021)\citenamefont
  {S.~Hermann}, \citenamefont {de~las Heras},\ and\ \citenamefont
  {Schmidt}}]{Hermann2021}%
  \BibitemOpen
  \bibfield  {author} {\bibinfo {author} {\bibfnamefont {S.}~\bibnamefont
  {S.~Hermann}}, \bibinfo {author} {\bibfnamefont {D.}~\bibnamefont {de~las
  Heras}}, \ and\ \bibinfo {author} {\bibfnamefont {M.}~\bibnamefont
  {Schmidt}},\ }\href@noop {} {\bibfield  {journal} {\bibinfo  {journal} {Mol.
  Phys.}\ }\textbf {\bibinfo {volume} {119}},\ \bibinfo {pages} {e1902585}
  (\bibinfo {year} {2021})}\BibitemShut {NoStop}%
\bibitem [{\citenamefont {Thutupalli}\ \emph {et~al.}(2011)\citenamefont
  {Thutupalli}, \citenamefont {Seemann},\ and\ \citenamefont
  {Herminghaus}}]{Thutupalli2011}%
  \BibitemOpen
  \bibfield  {author} {\bibinfo {author} {\bibfnamefont {S.}~\bibnamefont
  {Thutupalli}}, \bibinfo {author} {\bibfnamefont {R.}~\bibnamefont {Seemann}},
  \ and\ \bibinfo {author} {\bibfnamefont {S.}~\bibnamefont {Herminghaus}},\
  }\href@noop {} {\bibfield  {journal} {\bibinfo  {journal} {New Journal of
  Physics}\ }\textbf {\bibinfo {volume} {13}},\ \bibinfo {pages} {073021}
  (\bibinfo {year} {2011})}\BibitemShut {NoStop}%
\bibitem [{\citenamefont {D.~Nishiguchi}\ and\ \citenamefont
  {Sano}(2015)}]{Sano2015}%
  \BibitemOpen
  \bibfield  {author} {\bibinfo {author} {\bibfnamefont {D.}~\bibnamefont
  {D.~Nishiguchi}}\ and\ \bibinfo {author} {\bibfnamefont {M.}~\bibnamefont
  {Sano}},\ }\href@noop {} {\bibfield  {journal} {\bibinfo  {journal} {Physical
  Review E}\ }\textbf {\bibinfo {volume} {92}},\ \bibinfo {pages} {052309}
  (\bibinfo {year} {2015})}\BibitemShut {NoStop}%
\bibitem [{\citenamefont {Briand}\ and\ \citenamefont
  {Dauchot}(2016)}]{DauchotPRL2016}%
  \BibitemOpen
  \bibfield  {author} {\bibinfo {author} {\bibfnamefont {G.}~\bibnamefont
  {Briand}}\ and\ \bibinfo {author} {\bibfnamefont {O.}~\bibnamefont
  {Dauchot}},\ }\href@noop {} {\bibfield  {journal} {\bibinfo  {journal} {Phys.
  Rev. Lett.}\ }\textbf {\bibinfo {volume} {117}},\ \bibinfo {pages} {098004}
  (\bibinfo {year} {2016})}\BibitemShut {NoStop}%
\bibitem [{\citenamefont {Digregorio}\ \emph {et~al.}(2018)\citenamefont
  {Digregorio}, \citenamefont {Levis}, \citenamefont {Suma}, \citenamefont
  {Cugliandolo}, \citenamefont {Gonnella},\ and\ \citenamefont
  {Pagonabarraga}}]{DiGregorio_PRL_2018}%
  \BibitemOpen
  \bibfield  {author} {\bibinfo {author} {\bibfnamefont {P.}~\bibnamefont
  {Digregorio}}, \bibinfo {author} {\bibfnamefont {D.}~\bibnamefont {Levis}},
  \bibinfo {author} {\bibfnamefont {A.}~\bibnamefont {Suma}}, \bibinfo {author}
  {\bibfnamefont {L.~F.}\ \bibnamefont {Cugliandolo}}, \bibinfo {author}
  {\bibfnamefont {G.}~\bibnamefont {Gonnella}}, \ and\ \bibinfo {author}
  {\bibfnamefont {I.}~\bibnamefont {Pagonabarraga}},\ }\href {\doibase
  10.1103/PhysRevLett.121.098003} {\bibfield  {journal} {\bibinfo  {journal}
  {Phys. Rev. Lett.}\ }\textbf {\bibinfo {volume} {121}},\ \bibinfo {pages}
  {098003} (\bibinfo {year} {2018})}\BibitemShut {NoStop}%
\bibitem [{\citenamefont {Martin-Roca}\ \emph {et~al.}(2021)\citenamefont
  {Martin-Roca}, \citenamefont {Martinez}, \citenamefont {Alexander},
  \citenamefont {Diez}, \citenamefont {Aarts}, \citenamefont {Alarcon},
  \citenamefont {Ramírez},\ and\ \citenamefont {Valeriani}}]{JoseMIPS}%
  \BibitemOpen
  \bibfield  {author} {\bibinfo {author} {\bibfnamefont {J.}~\bibnamefont
  {Martin-Roca}}, \bibinfo {author} {\bibfnamefont {R.}~\bibnamefont
  {Martinez}}, \bibinfo {author} {\bibfnamefont {L.~C.}\ \bibnamefont
  {Alexander}}, \bibinfo {author} {\bibfnamefont {A.~L.}\ \bibnamefont {Diez}},
  \bibinfo {author} {\bibfnamefont {D.~G. A.~L.}\ \bibnamefont {Aarts}},
  \bibinfo {author} {\bibfnamefont {F.}~\bibnamefont {Alarcon}}, \bibinfo
  {author} {\bibfnamefont {J.}~\bibnamefont {Ramírez}}, \ and\ \bibinfo
  {author} {\bibfnamefont {C.}~\bibnamefont {Valeriani}},\ }\href {\doibase
  10.1063/5.0040141} {\bibfield  {journal} {\bibinfo  {journal} {The Journal of
  Chemical Physics}\ }\textbf {\bibinfo {volume} {154}},\ \bibinfo {pages}
  {164901} (\bibinfo {year} {2021})}\BibitemShut {NoStop}%
\bibitem [{\citenamefont {Solon}\ \emph {et~al.}(2018)\citenamefont {Solon},
  \citenamefont {Stenhammar}, \citenamefont {Cates}, \citenamefont {Kafri}, ,\
  and\ \citenamefont {Tailleur}}]{SolonNJP}%
  \BibitemOpen
  \bibfield  {author} {\bibinfo {author} {\bibfnamefont {A.~P.}\ \bibnamefont
  {Solon}}, \bibinfo {author} {\bibfnamefont {J.}~\bibnamefont {Stenhammar}},
  \bibinfo {author} {\bibfnamefont {M.~E.}\ \bibnamefont {Cates}}, \bibinfo
  {author} {\bibfnamefont {Y.}~\bibnamefont {Kafri}}, , \ and\ \bibinfo
  {author} {\bibfnamefont {J.}~\bibnamefont {Tailleur}},\ }\href@noop {}
  {\bibfield  {journal} {\bibinfo  {journal} {New J. Phys. .}\ }\textbf
  {\bibinfo {volume} {20}},\ \bibinfo {pages} {075001} (\bibinfo {year}
  {2018})}\BibitemShut {NoStop}%
\bibitem [{\citenamefont {Caprini}\ \emph {et~al.}(2019)\citenamefont
  {Caprini}, \citenamefont {Hernandez-Garcia}, \citenamefont {Lopez},\ and\
  \citenamefont {Marini Bettolo~Marconi}}]{Bettolo2019}%
  \BibitemOpen
  \bibfield  {author} {\bibinfo {author} {\bibfnamefont {L.}~\bibnamefont
  {Caprini}}, \bibinfo {author} {\bibfnamefont {E.}~\bibnamefont
  {Hernandez-Garcia}}, \bibinfo {author} {\bibfnamefont {C.}~\bibnamefont
  {Lopez}}, \ and\ \bibinfo {author} {\bibfnamefont {U.}~\bibnamefont {Marini
  Bettolo~Marconi}},\ }\href@noop {} {\bibfield  {journal} {\bibinfo  {journal}
  {Scientific Reports}\ }\textbf {\bibinfo {volume} {9}},\ \bibinfo {pages}
  {16687} (\bibinfo {year} {2019})}\BibitemShut {NoStop}%
\bibitem [{\citenamefont {Levis}\ \emph {et~al.}(2017)\citenamefont {Levis},
  \citenamefont {Codina},\ and\ \citenamefont
  {Pagonabarraga}}]{SoftMatterPagonabarragaa}%
  \BibitemOpen
  \bibfield  {author} {\bibinfo {author} {\bibfnamefont {D.}~\bibnamefont
  {Levis}}, \bibinfo {author} {\bibfnamefont {J.}~\bibnamefont {Codina}}, \
  and\ \bibinfo {author} {\bibfnamefont {I.}~\bibnamefont {Pagonabarraga}},\
  }\href@noop {} {\bibfield  {journal} {\bibinfo  {journal} {Soft Matter}\
  }\textbf {\bibinfo {volume} {13}},\ \bibinfo {pages} {8113} (\bibinfo {year}
  {2017})}\BibitemShut {NoStop}%
\bibitem [{\citenamefont {Shi}\ \emph {et~al.}(2019)\citenamefont {Shi},
  \citenamefont {Fausti}, \citenamefont {Chate}, \citenamefont {Nardini},\ and\
  \citenamefont {Solon}}]{SolonPRL2020}%
  \BibitemOpen
  \bibfield  {author} {\bibinfo {author} {\bibfnamefont {X.-q.}\ \bibnamefont
  {Shi}}, \bibinfo {author} {\bibfnamefont {G.}~\bibnamefont {Fausti}},
  \bibinfo {author} {\bibfnamefont {H.}~\bibnamefont {Chate}}, \bibinfo
  {author} {\bibfnamefont {C.}~\bibnamefont {Nardini}}, \ and\ \bibinfo
  {author} {\bibfnamefont {A.}~\bibnamefont {Solon}},\ }\href@noop {}
  {\bibfield  {journal} {\bibinfo  {journal} {Physical Review Letters}\
  }\textbf {\bibinfo {volume} {125}},\ \bibinfo {pages} {168001} (\bibinfo
  {year} {2019})}\BibitemShut {NoStop}%
\bibitem [{\citenamefont {Wysocki}\ \emph {et~al.}(2014)\citenamefont
  {Wysocki}, \citenamefont {Winkler},\ and\ \citenamefont
  {Gompper}}]{Wysocki2014}%
  \BibitemOpen
  \bibfield  {author} {\bibinfo {author} {\bibfnamefont {A.}~\bibnamefont
  {Wysocki}}, \bibinfo {author} {\bibfnamefont {R.}~\bibnamefont {Winkler}}, \
  and\ \bibinfo {author} {\bibfnamefont {G.}~\bibnamefont {Gompper}},\
  }\href@noop {} {\bibfield  {journal} {\bibinfo  {journal} {EPL}\ }\textbf
  {\bibinfo {volume} {105}},\ \bibinfo {pages} {48004} (\bibinfo {year}
  {2014})}\BibitemShut {NoStop}%
\bibitem [{\citenamefont {Lowen}(2020)}]{JCPLowen2020}%
  \BibitemOpen
  \bibfield  {author} {\bibinfo {author} {\bibfnamefont {H.}~\bibnamefont
  {Lowen}},\ }\href@noop {} {\bibfield  {journal} {\bibinfo  {journal} {J.
  Chem. Phys.}\ }\textbf {\bibinfo {volume} {152}},\ \bibinfo {pages} {040901}
  (\bibinfo {year} {2020})}\BibitemShut {NoStop}%
\bibitem [{\citenamefont {Su}\ \emph {et~al.}(2021)\citenamefont {Su},
  \citenamefont {Jiang},\ and\ \citenamefont {Hou}}]{Hou2021}%
  \BibitemOpen
  \bibfield  {author} {\bibinfo {author} {\bibfnamefont {J.}~\bibnamefont
  {Su}}, \bibinfo {author} {\bibfnamefont {H.}~\bibnamefont {Jiang}}, \ and\
  \bibinfo {author} {\bibfnamefont {Z.}~\bibnamefont {Hou}},\ }\href@noop {}
  {\bibfield  {journal} {\bibinfo  {journal} {New J. Phys.}\ }\textbf {\bibinfo
  {volume} {23}},\ \bibinfo {pages} {013005} (\bibinfo {year}
  {2021})}\BibitemShut {NoStop}%
\bibitem [{\citenamefont {Dittrich}\ \emph {et~al.}(2021)\citenamefont
  {Dittrich}, \citenamefont {Speck},\ and\ \citenamefont
  {Virnau}}]{MIPSvirnau}%
  \BibitemOpen
  \bibfield  {author} {\bibinfo {author} {\bibfnamefont {F.}~\bibnamefont
  {Dittrich}}, \bibinfo {author} {\bibfnamefont {T.}~\bibnamefont {Speck}}, \
  and\ \bibinfo {author} {\bibfnamefont {P.}~\bibnamefont {Virnau}},\
  }\href@noop {} {\bibfield  {journal} {\bibinfo  {journal} {Eur. Phys. J. E}\
  }\textbf {\bibinfo {volume} {44}},\ \bibinfo {pages} {53} (\bibinfo {year}
  {2021})}\BibitemShut {NoStop}%
\bibitem [{\citenamefont {Barriuso}\ \emph {et~al.}(2021)\citenamefont
  {Barriuso}, \citenamefont {Vanille}, \citenamefont {Alarcon}, \citenamefont
  {Pagonabarraga}, \citenamefont {Brito},\ and\ \citenamefont
  {Valeriani}}]{barri2021}%
  \BibitemOpen
  \bibfield  {author} {\bibinfo {author} {\bibfnamefont {C.~M.}\ \bibnamefont
  {Barriuso}}, \bibinfo {author} {\bibfnamefont {C.}~\bibnamefont {Vanille}},
  \bibinfo {author} {\bibfnamefont {F.}~\bibnamefont {Alarcon}}, \bibinfo
  {author} {\bibfnamefont {I.}~\bibnamefont {Pagonabarraga}}, \bibinfo {author}
  {\bibfnamefont {R.}~\bibnamefont {Brito}}, \ and\ \bibinfo {author}
  {\bibfnamefont {C.}~\bibnamefont {Valeriani}},\ }\href@noop {} {\enquote
  {\bibinfo {title} {Collective motion of run-and-tumble repulsive and
  attractive particles in one dimensional systems},}\ } (\bibinfo {year}
  {2021}),\ \Eprint {http://arxiv.org/abs/1912.01282} {arXiv:1912.01282
  [cond-mat.soft]} \BibitemShut {NoStop}%
\bibitem [{\citenamefont {Rogel~Rodriguez}\ \emph
  {et~al.}(2020{\natexlab{a}})\citenamefont {Rogel~Rodriguez}, \citenamefont
  {Alarcon}, \citenamefont {Martinez}, \citenamefont {Ramírez},\ and\
  \citenamefont {Valeriani}}]{DiegoMix2020}%
  \BibitemOpen
  \bibfield  {author} {\bibinfo {author} {\bibfnamefont {D.}~\bibnamefont
  {Rogel~Rodriguez}}, \bibinfo {author} {\bibfnamefont {F.}~\bibnamefont
  {Alarcon}}, \bibinfo {author} {\bibfnamefont {R.}~\bibnamefont {Martinez}},
  \bibinfo {author} {\bibfnamefont {J.}~\bibnamefont {Ramírez}}, \ and\
  \bibinfo {author} {\bibfnamefont {C.}~\bibnamefont {Valeriani}},\ }\href
  {\doibase 10.1039/C9SM01803D} {\bibfield  {journal} {\bibinfo  {journal}
  {Soft Matter}\ }\textbf {\bibinfo {volume} {16}},\ \bibinfo {pages} {1162}
  (\bibinfo {year} {2020}{\natexlab{a}})}\BibitemShut {NoStop}%
\bibitem [{\citenamefont {Stenhammar}\ \emph {et~al.}(2015)\citenamefont
  {Stenhammar}, \citenamefont {Wittkowski}, \citenamefont {Marenduzzo}, ,\ and\
  \citenamefont {Cates}}]{StenhammarPRL}%
  \BibitemOpen
  \bibfield  {author} {\bibinfo {author} {\bibfnamefont {J.}~\bibnamefont
  {Stenhammar}}, \bibinfo {author} {\bibfnamefont {R.}~\bibnamefont
  {Wittkowski}}, \bibinfo {author} {\bibfnamefont {D.}~\bibnamefont
  {Marenduzzo}}, , \ and\ \bibinfo {author} {\bibfnamefont {M.}~\bibnamefont
  {Cates}},\ }\href@noop {} {\bibfield  {journal} {\bibinfo  {journal}
  {Phys.Rev.Lett.}\ }\textbf {\bibinfo {volume} {114}},\ \bibinfo {pages}
  {018301} (\bibinfo {year} {2015})}\BibitemShut {NoStop}%
\bibitem [{\citenamefont {Wysocki}\ \emph {et~al.}(2016)\citenamefont
  {Wysocki}, \citenamefont {Winkler},\ and\ \citenamefont
  {Gompper}}]{Wysocki2016}%
  \BibitemOpen
  \bibfield  {author} {\bibinfo {author} {\bibfnamefont {A.}~\bibnamefont
  {Wysocki}}, \bibinfo {author} {\bibfnamefont {R.}~\bibnamefont {Winkler}}, \
  and\ \bibinfo {author} {\bibfnamefont {G.}~\bibnamefont {Gompper}},\
  }\href@noop {} {\bibfield  {journal} {\bibinfo  {journal} {New J. Phys.}\
  }\textbf {\bibinfo {volume} {18}},\ \bibinfo {pages} {123030} (\bibinfo
  {year} {2016})}\BibitemShut {NoStop}%
\bibitem [{\citenamefont {Siebert}\ \emph {et~al.}(2018)\citenamefont
  {Siebert}, \citenamefont {Dittrich}, \citenamefont {Schmid}, \citenamefont
  {Binder}, \citenamefont {Speck},\ and\ \citenamefont
  {Virnau}}]{MIPSBand_BinderCum}%
  \BibitemOpen
  \bibfield  {author} {\bibinfo {author} {\bibfnamefont {J.~T.}\ \bibnamefont
  {Siebert}}, \bibinfo {author} {\bibfnamefont {F.}~\bibnamefont {Dittrich}},
  \bibinfo {author} {\bibfnamefont {F.}~\bibnamefont {Schmid}}, \bibinfo
  {author} {\bibfnamefont {K.}~\bibnamefont {Binder}}, \bibinfo {author}
  {\bibfnamefont {T.}~\bibnamefont {Speck}}, \ and\ \bibinfo {author}
  {\bibfnamefont {P.}~\bibnamefont {Virnau}},\ }\href {\doibase
  10.1103/PhysRevE.98.030601} {\bibfield  {journal} {\bibinfo  {journal} {Phys.
  Rev. E}\ }\textbf {\bibinfo {volume} {98}},\ \bibinfo {pages} {030601}
  (\bibinfo {year} {2018})}\BibitemShut {NoStop}%
\bibitem [{\citenamefont {Patch}\ \emph {et~al.}(2018)\citenamefont {Patch},
  \citenamefont {Sussman}, \citenamefont {Yllanes},\ and\ \citenamefont
  {Marchetti}}]{Patch_SoftMat_2018}%
  \BibitemOpen
  \bibfield  {author} {\bibinfo {author} {\bibfnamefont {A.}~\bibnamefont
  {Patch}}, \bibinfo {author} {\bibfnamefont {D.~M.}\ \bibnamefont {Sussman}},
  \bibinfo {author} {\bibfnamefont {D.}~\bibnamefont {Yllanes}}, \ and\
  \bibinfo {author} {\bibfnamefont {M.~C.}\ \bibnamefont {Marchetti}},\ }\href
  {\doibase 10.1039/C8SM00899J} {\bibfield  {journal} {\bibinfo  {journal}
  {Soft Matter}\ }\textbf {\bibinfo {volume} {14}},\ \bibinfo {pages} {7435}
  (\bibinfo {year} {2018})}\BibitemShut {NoStop}%
\bibitem [{\citenamefont {Paliwal}\ \emph {et~al.}(2017)\citenamefont
  {Paliwal}, \citenamefont {Prymidis}, \citenamefont {Filion},\ and\
  \citenamefont {Dijkstra}}]{JCPdijkstra2017}%
  \BibitemOpen
  \bibfield  {author} {\bibinfo {author} {\bibfnamefont {S.}~\bibnamefont
  {Paliwal}}, \bibinfo {author} {\bibfnamefont {V.}~\bibnamefont {Prymidis}},
  \bibinfo {author} {\bibfnamefont {L.}~\bibnamefont {Filion}}, \ and\ \bibinfo
  {author} {\bibfnamefont {M.}~\bibnamefont {Dijkstra}},\ }\href@noop {}
  {\bibfield  {journal} {\bibinfo  {journal} {J. Chem. Phys.}\ }\textbf
  {\bibinfo {volume} {147}},\ \bibinfo {pages} {084902} (\bibinfo {year}
  {2017})}\BibitemShut {NoStop}%
\bibitem [{\citenamefont {Fausti}\ \emph {et~al.}(2021)\citenamefont {Fausti},
  \citenamefont {Tjhung}, \citenamefont {Cates},\ and\ \citenamefont
  {Nardini}}]{Nardini_2021}%
  \BibitemOpen
  \bibfield  {author} {\bibinfo {author} {\bibfnamefont {G.}~\bibnamefont
  {Fausti}}, \bibinfo {author} {\bibfnamefont {E.}~\bibnamefont {Tjhung}},
  \bibinfo {author} {\bibfnamefont {M.~E.}\ \bibnamefont {Cates}}, \ and\
  \bibinfo {author} {\bibfnamefont {C.}~\bibnamefont {Nardini}},\ }\href
  {\doibase 10.1103/PhysRevLett.127.068001} {\bibfield  {journal} {\bibinfo
  {journal} {Phys. Rev. Lett.}\ }\textbf {\bibinfo {volume} {127}},\ \bibinfo
  {pages} {068001} (\bibinfo {year} {2021})}\BibitemShut {NoStop}%
\bibitem [{\citenamefont {F}\ \emph {et~al.}(2015)\citenamefont {F},
  \citenamefont {Theurkauff}, \citenamefont {Levis}, \citenamefont {Ybert},
  \citenamefont {Bocquet}, \citenamefont {Berthier},\ and\ \citenamefont
  {Cottin-Bizonne}}]{Ginot2015}%
  \BibitemOpen
  \bibfield  {author} {\bibinfo {author} {\bibfnamefont {G.}~\bibnamefont {F}},
  \bibinfo {author} {\bibfnamefont {I.}~\bibnamefont {Theurkauff}}, \bibinfo
  {author} {\bibfnamefont {D.}~\bibnamefont {Levis}}, \bibinfo {author}
  {\bibfnamefont {C.}~\bibnamefont {Ybert}}, \bibinfo {author} {\bibfnamefont
  {L.}~\bibnamefont {Bocquet}}, \bibinfo {author} {\bibfnamefont
  {L.}~\bibnamefont {Berthier}}, \ and\ \bibinfo {author} {\bibfnamefont
  {C.}~\bibnamefont {Cottin-Bizonne}},\ }\href@noop {} {\bibfield  {journal}
  {\bibinfo  {journal} {Phys. Rev. X}\ }\textbf {\bibinfo {volume} {5}},\
  \bibinfo {pages} {011004} (\bibinfo {year} {2015})}\BibitemShut {NoStop}%
\bibitem [{\citenamefont {Fily}\ \emph {et~al.}(2014)\citenamefont {Fily},
  \citenamefont {Baskaran},\ and\ \citenamefont {Hagan}}]{Fily2014}%
  \BibitemOpen
  \bibfield  {author} {\bibinfo {author} {\bibfnamefont {Y.}~\bibnamefont
  {Fily}}, \bibinfo {author} {\bibfnamefont {A.}~\bibnamefont {Baskaran}}, \
  and\ \bibinfo {author} {\bibfnamefont {M.~F.}\ \bibnamefont {Hagan}},\
  }\href@noop {} {\bibfield  {journal} {\bibinfo  {journal} {Soft matter}\
  }\textbf {\bibinfo {volume} {10}},\ \bibinfo {pages} {5609} (\bibinfo {year}
  {2014})}\BibitemShut {NoStop}%
\bibitem [{\citenamefont {Takatori}\ \emph {et~al.}(2014)\citenamefont
  {Takatori}, \citenamefont {Yan},\ and\ \citenamefont {Brady}}]{Takatori2014}%
  \BibitemOpen
  \bibfield  {author} {\bibinfo {author} {\bibfnamefont {S.~C.}\ \bibnamefont
  {Takatori}}, \bibinfo {author} {\bibfnamefont {W.}~\bibnamefont {Yan}}, \
  and\ \bibinfo {author} {\bibfnamefont {J.~F.}\ \bibnamefont {Brady}},\
  }\href@noop {} {\bibfield  {journal} {\bibinfo  {journal} {Phys. Rev. Lett.}\
  }\textbf {\bibinfo {volume} {113}},\ \bibinfo {pages} {1} (\bibinfo {year}
  {2014})}\BibitemShut {NoStop}%
\bibitem [{\citenamefont {Mallory}\ \emph {et~al.}(2020)\citenamefont
  {Mallory}, \citenamefont {Omar},\ and\ \citenamefont {Brady}}]{mallory2020}%
  \BibitemOpen
  \bibfield  {author} {\bibinfo {author} {\bibfnamefont {S.~A.}\ \bibnamefont
  {Mallory}}, \bibinfo {author} {\bibfnamefont {A.~K.}\ \bibnamefont {Omar}}, \
  and\ \bibinfo {author} {\bibfnamefont {J.~F.}\ \bibnamefont {Brady}},\
  }\href@noop {} {\enquote {\bibinfo {title} {Dynamic overlap concentration
  scale of active colloids},}\ } (\bibinfo {year} {2020}),\ \Eprint
  {http://arxiv.org/abs/2009.06092} {arXiv:2009.06092 [cond-mat.soft]}
  \BibitemShut {NoStop}%
\bibitem [{\citenamefont {Epstein}\ \emph {et~al.}(2019)\citenamefont
  {Epstein}, \citenamefont {Klymko},\ and\ \citenamefont
  {Mandadapu}}]{Epstein2019}%
  \BibitemOpen
  \bibfield  {author} {\bibinfo {author} {\bibfnamefont {J.~M.}\ \bibnamefont
  {Epstein}}, \bibinfo {author} {\bibfnamefont {K.}~\bibnamefont {Klymko}}, \
  and\ \bibinfo {author} {\bibfnamefont {K.~K.}\ \bibnamefont {Mandadapu}},\
  }\href@noop {} {\bibfield  {journal} {\bibinfo  {journal} {J. Chem. Phys.}\
  }\textbf {\bibinfo {volume} {150}},\ \bibinfo {pages} {1} (\bibinfo {year}
  {2019})}\BibitemShut {NoStop}%
\bibitem [{\citenamefont {Winkler}\ \emph {et~al.}(2015)\citenamefont
  {Winkler}, \citenamefont {Wysocki},\ and\ \citenamefont
  {Gompper}}]{Winkler2015}%
  \BibitemOpen
  \bibfield  {author} {\bibinfo {author} {\bibfnamefont {R.~G.}\ \bibnamefont
  {Winkler}}, \bibinfo {author} {\bibfnamefont {A.}~\bibnamefont {Wysocki}}, \
  and\ \bibinfo {author} {\bibfnamefont {G.}~\bibnamefont {Gompper}},\
  }\href@noop {} {\bibfield  {journal} {\bibinfo  {journal} {Soft Matter}\
  }\textbf {\bibinfo {volume} {11}},\ \bibinfo {pages} {6680} (\bibinfo {year}
  {2015})}\BibitemShut {NoStop}%
\bibitem [{\citenamefont {Patch}\ \emph {et~al.}(2017)\citenamefont {Patch},
  \citenamefont {Yllanes},\ and\ \citenamefont {Marchetti}}]{Patch2017}%
  \BibitemOpen
  \bibfield  {author} {\bibinfo {author} {\bibfnamefont {A.}~\bibnamefont
  {Patch}}, \bibinfo {author} {\bibfnamefont {D.}~\bibnamefont {Yllanes}}, \
  and\ \bibinfo {author} {\bibfnamefont {M.~C.}\ \bibnamefont {Marchetti}},\
  }\href@noop {} {\bibfield  {journal} {\bibinfo  {journal} {Phys. Rev. E}\
  }\textbf {\bibinfo {volume} {95}},\ \bibinfo {pages} {012601} (\bibinfo
  {year} {2017})}\BibitemShut {NoStop}%
\bibitem [{\citenamefont {Solon}\ \emph {et~al.}(2015)\citenamefont {Solon},
  \citenamefont {Fily}, \citenamefont {Baskaran}, \citenamefont {Cates},
  \citenamefont {Kafri}, \citenamefont {Kardar},\ and\ \citenamefont
  {Tailleur}}]{SolonNatPhys}%
  \BibitemOpen
  \bibfield  {author} {\bibinfo {author} {\bibfnamefont {A.}~\bibnamefont
  {Solon}}, \bibinfo {author} {\bibfnamefont {Y.}~\bibnamefont {Fily}},
  \bibinfo {author} {\bibfnamefont {A.}~\bibnamefont {Baskaran}}, \bibinfo
  {author} {\bibfnamefont {M.}~\bibnamefont {Cates}}, \bibinfo {author}
  {\bibfnamefont {Y.}~\bibnamefont {Kafri}}, \bibinfo {author} {\bibfnamefont
  {M.}~\bibnamefont {Kardar}}, \ and\ \bibinfo {author} {\bibfnamefont
  {J.}~\bibnamefont {Tailleur}},\ }\href@noop {} {\bibfield  {journal}
  {\bibinfo  {journal} {Nat. Phys.}\ }\textbf {\bibinfo {volume} {11}},\
  \bibinfo {pages} {673} (\bibinfo {year} {2015})}\BibitemShut {NoStop}%
\bibitem [{\citenamefont {Speck}\ and\ \citenamefont {Jack}(2016)}]{Speck2016}%
  \BibitemOpen
  \bibfield  {author} {\bibinfo {author} {\bibfnamefont {T.}~\bibnamefont
  {Speck}}\ and\ \bibinfo {author} {\bibfnamefont {R.}~\bibnamefont {Jack}},\
  }\href@noop {} {\bibfield  {journal} {\bibinfo  {journal} {Phys. Rev. E}\
  }\textbf {\bibinfo {volume} {93}},\ \bibinfo {pages} {062605} (\bibinfo
  {year} {2016})}\BibitemShut {NoStop}%
\bibitem [{\citenamefont {Fily}\ \emph {et~al.}(2018)\citenamefont {Fily},
  \citenamefont {Kafri}, \citenamefont {Solon}, \citenamefont {Tailleur},\ and\
  \citenamefont {Turner}}]{Fily2018}%
  \BibitemOpen
  \bibfield  {author} {\bibinfo {author} {\bibfnamefont {Y.}~\bibnamefont
  {Fily}}, \bibinfo {author} {\bibfnamefont {Y.}~\bibnamefont {Kafri}},
  \bibinfo {author} {\bibfnamefont {A.}~\bibnamefont {Solon}}, \bibinfo
  {author} {\bibfnamefont {J.}~\bibnamefont {Tailleur}}, \ and\ \bibinfo
  {author} {\bibfnamefont {A.}~\bibnamefont {Turner}},\ }\href@noop {}
  {\bibfield  {journal} {\bibinfo  {journal} {J. Phys. A}\ }\textbf {\bibinfo
  {volume} {51}},\ \bibinfo {pages} {044003} (\bibinfo {year}
  {2018})}\BibitemShut {NoStop}%
\bibitem [{\citenamefont {Das}\ \emph {et~al.}(2019)\citenamefont {Das},
  \citenamefont {Gompper},\ and\ \citenamefont {Winkler}}]{Das2019}%
  \BibitemOpen
  \bibfield  {author} {\bibinfo {author} {\bibfnamefont {S.}~\bibnamefont
  {Das}}, \bibinfo {author} {\bibfnamefont {G.}~\bibnamefont {Gompper}}, \ and\
  \bibinfo {author} {\bibfnamefont {R.}~\bibnamefont {Winkler}},\ }\href@noop
  {} {\bibfield  {journal} {\bibinfo  {journal} {Sci. Rep.}\ }\textbf {\bibinfo
  {volume} {9}},\ \bibinfo {pages} {6608} (\bibinfo {year} {2019})}\BibitemShut
  {NoStop}%
\bibitem [{\citenamefont {Trokhymchuk}\ and\ \citenamefont
  {Alejandre}(2013)}]{Alejandre1999}%
  \BibitemOpen
  \bibfield  {author} {\bibinfo {author} {\bibfnamefont {A.}~\bibnamefont
  {Trokhymchuk}}\ and\ \bibinfo {author} {\bibfnamefont {J.}~\bibnamefont
  {Alejandre}},\ }\href@noop {} {\bibfield  {journal} {\bibinfo  {journal} {J.
  Chem. Lett.}\ }\textbf {\bibinfo {volume} {111}},\ \bibinfo {pages} {8510}
  (\bibinfo {year} {2013})}\BibitemShut {NoStop}%
\bibitem [{\citenamefont {Lauersdorf}\ \emph {et~al.}(2021)\citenamefont
  {Lauersdorf}, \citenamefont {Kolb}, \citenamefont {Moradi}, \citenamefont
  {E},\ and\ \citenamefont {Klotsa}}]{SoftMatterKlotsa}%
  \BibitemOpen
  \bibfield  {author} {\bibinfo {author} {\bibfnamefont {N.}~\bibnamefont
  {Lauersdorf}}, \bibinfo {author} {\bibfnamefont {T.}~\bibnamefont {Kolb}},
  \bibinfo {author} {\bibfnamefont {M.}~\bibnamefont {Moradi}}, \bibinfo
  {author} {\bibfnamefont {E.~N.}\ \bibnamefont {E}}, \ and\ \bibinfo {author}
  {\bibfnamefont {D.}~\bibnamefont {Klotsa}},\ }\href@noop {} {\bibfield
  {journal} {\bibinfo  {journal} {Soft Matter}\ }\textbf {\bibinfo {volume}
  {17}},\ \bibinfo {pages} {6337} (\bibinfo {year} {2021})}\BibitemShut
  {NoStop}%
\bibitem [{\citenamefont {Rowlinson}\ and\ \citenamefont
  {Widom}(1982)}]{Rowlinson1982}%
  \BibitemOpen
  \bibfield  {author} {\bibinfo {author} {\bibfnamefont {J.~S.}\ \bibnamefont
  {Rowlinson}}\ and\ \bibinfo {author} {\bibfnamefont {B.}~\bibnamefont
  {Widom}},\ }\href@noop {} {\emph {\bibinfo {title} {Molecular Theory of
  Capillarity}}}\ (\bibinfo  {publisher} {Claredon Press},\ \bibinfo {year}
  {1982})\BibitemShut {NoStop}%
\bibitem [{\citenamefont {Buff}\ \emph {et~al.}(1965)\citenamefont {Buff},
  \citenamefont {Lovett},\ and\ \citenamefont {Stillinger}}]{Buff_CWT_1965}%
  \BibitemOpen
  \bibfield  {author} {\bibinfo {author} {\bibfnamefont {F.~P.}\ \bibnamefont
  {Buff}}, \bibinfo {author} {\bibfnamefont {R.~A.}\ \bibnamefont {Lovett}}, \
  and\ \bibinfo {author} {\bibfnamefont {F.~H.}\ \bibnamefont {Stillinger}},\
  }\href@noop {} {\bibfield  {journal} {\bibinfo  {journal} {Phys. Rev. Lett.}\
  }\textbf {\bibinfo {volume} {15}},\ \bibinfo {pages} {621} (\bibinfo {year}
  {1965})}\BibitemShut {NoStop}%
\bibitem [{\citenamefont {Chac\'{o}n}\ and\ \citenamefont
  {Tarazona}(2003)}]{chacon_ISM_2003_PRL}%
  \BibitemOpen
  \bibfield  {author} {\bibinfo {author} {\bibfnamefont {E.}~\bibnamefont
  {Chac\'{o}n}}\ and\ \bibinfo {author} {\bibfnamefont {P.}~\bibnamefont
  {Tarazona}},\ }\href@noop {} {\bibfield  {journal} {\bibinfo  {journal}
  {Phys. Rev. Lett.}\ }\textbf {\bibinfo {volume} {91}},\ \bibinfo {pages}
  {166103} (\bibinfo {year} {2003})}\BibitemShut {NoStop}%
\bibitem [{\citenamefont {Delgado-Buscalioni}\ \emph
  {et~al.}(2008)\citenamefont {Delgado-Buscalioni}, \citenamefont
  {Chac\'{o}n},\ and\ \citenamefont {Tarazona}}]{HIDRODYNAMIC_ISM_PRL_2008}%
  \BibitemOpen
  \bibfield  {author} {\bibinfo {author} {\bibfnamefont {R.}~\bibnamefont
  {Delgado-Buscalioni}}, \bibinfo {author} {\bibfnamefont {E.}~\bibnamefont
  {Chac\'{o}n}}, \ and\ \bibinfo {author} {\bibfnamefont {P.}~\bibnamefont
  {Tarazona}},\ }\href@noop {} {\bibfield  {journal} {\bibinfo  {journal}
  {Phys. Rev. Lett.}\ }\textbf {\bibinfo {volume} {101}},\ \bibinfo {pages}
  {106102} (\bibinfo {year} {2008})}\BibitemShut {NoStop}%
\bibitem [{\citenamefont {Tarazona}\ \emph {et~al.}(2007)\citenamefont
  {Tarazona}, \citenamefont {Checa},\ and\ \citenamefont
  {Chac\'on}}]{Tarazona_CW_2007_PRL}%
  \BibitemOpen
  \bibfield  {author} {\bibinfo {author} {\bibfnamefont {P.}~\bibnamefont
  {Tarazona}}, \bibinfo {author} {\bibfnamefont {R.}~\bibnamefont {Checa}}, \
  and\ \bibinfo {author} {\bibfnamefont {E.}~\bibnamefont {Chac\'on}},\
  }\href@noop {} {\bibfield  {journal} {\bibinfo  {journal} {Phys. Rev. Lett.}\
  }\textbf {\bibinfo {volume} {99}},\ \bibinfo {pages} {196101} (\bibinfo
  {year} {2007})}\BibitemShut {NoStop}%
\bibitem [{\citenamefont {Plimpton}(1995)}]{LAMMPS}%
  \BibitemOpen
  \bibfield  {author} {\bibinfo {author} {\bibfnamefont {S.}~\bibnamefont
  {Plimpton}},\ }\href@noop {} {\bibfield  {journal} {\bibinfo  {journal} {J.
  Comp. Phys.}\ }\textbf {\bibinfo {volume} {117}},\ \bibinfo {pages} {1}
  (\bibinfo {year} {1995})}\BibitemShut {NoStop}%
\bibitem [{\citenamefont {Rogel~Rodriguez}\ \emph
  {et~al.}(2020{\natexlab{b}})\citenamefont {Rogel~Rodriguez}, \citenamefont
  {Alarcon}, \citenamefont {Martinez}, \citenamefont {Ramírez},\ and\
  \citenamefont {Valeriani}}]{Diego_SoftMat2020}%
  \BibitemOpen
  \bibfield  {author} {\bibinfo {author} {\bibfnamefont {D.}~\bibnamefont
  {Rogel~Rodriguez}}, \bibinfo {author} {\bibfnamefont {F.}~\bibnamefont
  {Alarcon}}, \bibinfo {author} {\bibfnamefont {R.}~\bibnamefont {Martinez}},
  \bibinfo {author} {\bibfnamefont {J.}~\bibnamefont {Ramírez}}, \ and\
  \bibinfo {author} {\bibfnamefont {C.}~\bibnamefont {Valeriani}},\ }\href
  {\doibase 10.1039/C9SM01803D} {\bibfield  {journal} {\bibinfo  {journal}
  {Soft Matter}\ }\textbf {\bibinfo {volume} {16}},\ \bibinfo {pages} {1162}
  (\bibinfo {year} {2020}{\natexlab{b}})}\BibitemShut {NoStop}%
\bibitem [{Note1()}]{Note1}%
  \BibitemOpen
  \bibinfo {note} {Note that this ratio is related to the $\xi $ reported in
  Ref\cite {AlarconSoftmatter}}\BibitemShut {NoStop}%
\bibitem [{\citenamefont {Caporusso}\ \emph {et~al.}(2020)\citenamefont
  {Caporusso}, \citenamefont {Digregorio}, \citenamefont {Levis}, \citenamefont
  {Cugliandolo},\ and\ \citenamefont {Gonnella}}]{Caporusso_PRL_2020}%
  \BibitemOpen
  \bibfield  {author} {\bibinfo {author} {\bibfnamefont {C.~B.}\ \bibnamefont
  {Caporusso}}, \bibinfo {author} {\bibfnamefont {P.}~\bibnamefont
  {Digregorio}}, \bibinfo {author} {\bibfnamefont {D.}~\bibnamefont {Levis}},
  \bibinfo {author} {\bibfnamefont {L.~F.}\ \bibnamefont {Cugliandolo}}, \ and\
  \bibinfo {author} {\bibfnamefont {G.}~\bibnamefont {Gonnella}},\ }\href@noop
  {} {\bibfield  {journal} {\bibinfo  {journal} {Phys. Rev. Lett.}\ }\textbf
  {\bibinfo {volume} {125}},\ \bibinfo {pages} {178004} (\bibinfo {year}
  {2020})}\BibitemShut {NoStop}%
\bibitem [{\citenamefont {Digregorio}\ \emph {et~al.}(2021)\citenamefont
  {Digregorio}, \citenamefont {Levis}, \citenamefont {Cugliandolo},
  \citenamefont {Gonnella},\ and\ \citenamefont
  {Pagonabarraga}}]{digregorio_arXiv_2021}%
  \BibitemOpen
  \bibfield  {author} {\bibinfo {author} {\bibfnamefont {P.}~\bibnamefont
  {Digregorio}}, \bibinfo {author} {\bibfnamefont {D.}~\bibnamefont {Levis}},
  \bibinfo {author} {\bibfnamefont {L.~F.}\ \bibnamefont {Cugliandolo}},
  \bibinfo {author} {\bibfnamefont {G.}~\bibnamefont {Gonnella}}, \ and\
  \bibinfo {author} {\bibfnamefont {I.}~\bibnamefont {Pagonabarraga}},\
  }\href@noop {} {\enquote {\bibinfo {title} {Clustering of topological defects
  in two-dimensional melting of active and passive disks},}\ } (\bibinfo {year}
  {2021}),\ \Eprint {http://arxiv.org/abs/1911.06366} {arXiv:1911.06366
  [cond-mat.soft]} \BibitemShut {NoStop}%
\bibitem [{\citenamefont {Paliwal}\ and\ \citenamefont
  {Dijkstra}(2020)}]{Paliwal_PhysRevResearch_2020}%
  \BibitemOpen
  \bibfield  {author} {\bibinfo {author} {\bibfnamefont {S.}~\bibnamefont
  {Paliwal}}\ and\ \bibinfo {author} {\bibfnamefont {M.}~\bibnamefont
  {Dijkstra}},\ }\href {\doibase 10.1103/PhysRevResearch.2.012013} {\bibfield
  {journal} {\bibinfo  {journal} {Phys. Rev. Research}\ }\textbf {\bibinfo
  {volume} {2}},\ \bibinfo {pages} {012013} (\bibinfo {year}
  {2020})}\BibitemShut {NoStop}%
\bibitem [{\citenamefont {Klamser}\ \emph {et~al.}(2018)\citenamefont
  {Klamser}, \citenamefont {S.C.},\ and\ \citenamefont
  {Krauth}}]{Kamser_NatCom_2018}%
  \BibitemOpen
  \bibfield  {author} {\bibinfo {author} {\bibfnamefont {J.}~\bibnamefont
  {Klamser}}, \bibinfo {author} {\bibfnamefont {S.~K.}\ \bibnamefont {S.C.}}, \
  and\ \bibinfo {author} {\bibfnamefont {W.}~\bibnamefont {Krauth}},\ }\href
  {\doibase 10.1038/s41467-018-07491-5} {\bibfield  {journal} {\bibinfo
  {journal} {Phys. Rev. Lett.}\ }\textbf {\bibinfo {volume} {9}},\ \bibinfo
  {pages} {5045} (\bibinfo {year} {2018})}\BibitemShut {NoStop}%
\end{thebibliography}
%Control: key (0)
%Control: author (8) initials jnrlst
%Control: editor formatted (1) identically to author
%Control: production of article title (-1) disabled
%Control: page (0) single
%Control: year (1) truncated
%Control: production of eprint (0) enabled
\providecommand{\noopsort}[1]{}\providecommand{\singleletter}[1]{#1}%

\end{document}